\documentclass{article}

\usepackage{microtype}
\usepackage{graphicx}
\usepackage{subfigure}
\usepackage{booktabs} 
\usepackage{hyperref}

\usepackage{amsmath}
\usepackage{amssymb}

\usepackage[accepted]{icml2019}
\icmltitlerunning{Generalized Approximate Survey Propagation}

\DeclareMathOperator{\dd}{\mathrm{d}}

\DeclareMathOperator{\sign}{sign}

\begin{document}
\newcommand{\bx}{\boldsymbol{x}}
\newcommand{\by}{\boldsymbol{y}}
\newcommand{\bF}{\boldsymbol{F}}
\newcommand{\bw}{\boldsymbol{\omega}}
\newcommand{\bet}{\boldsymbol{\eta}}
\newcommand{\bB}{\boldsymbol{B}}

\twocolumn[
\icmltitle{Generalized Approximate Survey Propagation \\
          for High-Dimensional Estimation}



\begin{icmlauthorlist}

\icmlauthor{Luca Saglietti}{microsoft,iigm}
\icmlauthor{Yue M. Lu}{harvard}
\icmlauthor{Carlo Lucibello}{bidsa,microsoft}
\end{icmlauthorlist}

\icmlaffiliation{iigm}{Italian Institute for Genomic Medicine, Turin, Italy}
\icmlaffiliation{microsoft}{Microsoft Research New England, Cambridge, MA 02142, USA}
\icmlaffiliation{harvard}{John A. Paulson School of Engineering and Applied Sciences, Harvard University, Cambridge, MA 02138, USA}
\icmlaffiliation{bidsa}{Bocconi Institute for Data Science and Analytics, Bocconi University, Milan, Italy}

\icmlcorrespondingauthor{Carlo Lucibello}{carlo.lucibello@unibocconi.it}
\icmlcorrespondingauthor{Luca Saglietti}{luca.saglietti@gmail.com}

\icmlkeywords{Machine Learning, ICML, Signal Processing, Message Passing}
\vskip 0.3in] 

\printAffiliationsAndNotice{}  

\begin{abstract}
In Generalized Linear Estimation (GLE) problems, we seek to estimate a signal that is observed through a linear transform followed by a component-wise, possibly nonlinear and noisy, channel. In the Bayesian optimal setting, Generalized Approximate Message Passing (GAMP) is known to achieve optimal performance for GLE. However, its performance can significantly degrade whenever there is a mismatch between the assumed and the true generative model, a situation frequently encountered in practice. In this paper, we propose a new algorithm, named Generalized Approximate Survey Propagation (GASP), for solving GLE in the presence of prior or model mis-specifications. As a prototypical example, we consider the phase retrieval problem, where we show that GASP outperforms the corresponding GAMP, reducing the reconstruction threshold and, for certain choices of its parameters, approaching Bayesian optimal performance. Furthermore, we present a set of State Evolution equations that exactly characterize the dynamics of GASP in the high-dimensional limit.
\end{abstract}

\section{Introduction}
\label{sec:intro}

\emph{Approximate message
passing} (AMP) algorithms have become a well established tool in the
study of inference problems \cite{donoho2009message,donoho2016high,advani2016statistical} that can be represented by dense graphical models. An important feature of AMP is that its dynamical behavior in the
large system limit can be exactly predicted through a dynamical system involving only scalar quantities called \emph{State
Evolution} (SE) \cite{bayati2011dynamics}. This relationship paved the way for a series of rigorous
results \cite{rangan2012iterative,deshpande2014information,deshpande2016asymptotic}. It also helps clarify the connection to several fascinating predictions obtained through the \emph{replica analysis} in statistical physics \cite{mezard1987spin}. In the optimal Bayesian setting, where one has perfect information on the process underlying data generation, AMP has been empirically shown to achieve  optimal performances among polynomial algorithms for many different problems. However, in the more realistic case of mismatch between the assumed and the true generative model, i.e. when AMP is not derived on the true posterior distribution, it may become sub-optimal. A possible source of problems for the AMP class of algorithms is the outbreak of \emph{Replica Symmetry Breaking} \cite{mezard1987spin}, a scenario where an exponential number of fixed point and algorithmic barriers dominate the free energy landscape explored by AMP. This phenomena can be accentuated in case of model mismatch: a notable example is maximum likelihood estimation (as opposed to estimation by the posterior mean, which corresponds to the low temperature limit of a statistical physics model.

These considerations  are well known within the physics community of disordered systems \cite{krzakala2016statistical}, where the problem of signal estimation is informally referred to as ``crystal hunting''. Estimation problems in high dimensions are characterized by a complex energy-entropy competition where the true signal is hidden in a vast and potentially rough landscape. In a wide class of problems, one observes the
presence of a algorithmically ``\emph{hard}'' phase for some range of values for the parameters defining the problem (e.g. signal-to-noise ration). In this regime, all known polynomial complexity algorithms fail to saturate the information theoretic bound \cite{ricci2019typology}. While reconstruction is possible in principle, algorithms are trapped in a region of the configuration space with low overlap with the signal and many local minima \cite{antenucci2019glassy,ros2019complex}.

In a recent work \cite{antenucci2019approximate}, 
a novel message-passing algorithm, \emph{Approximate Survey Propagation} (ASP), was introduced in the context of low-rank. The algorithm is based on the \emph{1-step Replica Symmetry Breaking} (1RSB) ansatz from spin glass theory \cite{mezard1987spin}, which was specifically developed to deal with landscapes populated by exponentially  many local minima. It was shown that ASP on the mismatched model could reach the performance  of (but not improve on) matched AMP and do far better than mismatched AMP \cite{antenucci2019glassy,antenucci2019approximate}.
In the present paper, we build upon these previous works and derive
the ASP algorithm for \emph{Generalized Linear Estimation} (GLE) models.
Since the extension of AMP to GLE problems is commonly known as GAMP,
we call \emph{Generalized Approximate Survey Propagation} (GASP), our extension of ASP. We will show that also in this case, in presence of model mismatch, (G)ASP improves over the corresponding (G)AMP. 

\section{Model specification}

An instance of the general class of models to which GASP can be applied is defined, for some integer $M$ and $N$, by an \emph{observed signal} $\by\in \mathbb{R}^M$ 
and an $M\times N$ \emph{observation matrix}  $\bF$. Clearly, this scenario encompasses also GLE. We denote with $\bF^\mu,\ \mu \in [M]$, the rows of $\bF$ and refer to the ratio $\alpha= M/N$ as the \emph{sampling ratio}  of $\bF$.
We consider a probability density distribution $p(\bx)$ on a (possibly discrete) space $\chi^N$, $\chi\subseteq \mathbb{R}$, defined as:
\begin{equation}
p(\bx) = \frac{1}{Z}\  e^{-\beta \mathcal{H}_{\by,\bF}(\bx)}, 
\label{eq:p}
\end{equation}
where, following statistical physics jargon, $\beta$ plays the role of an inverse temperature, $Z$ is a normalization factor called partition function (both $Z$ and $p$ implicitly depend on $\beta,\by$ and $\bF$),  and $\mathcal{H}_{\by,\bF}$ is the Hamiltonian of the model, that in our setting takes the form: 
\begin{equation}
\mathcal{H}_{\by,\bF}(\bx)=\sum_{\mu=1}^{M}\ell\big(y_{\mu},\,\langle\bF^{\mu},\bx\rangle\big)+ \sum_{i=1}^{N}r(x_{i}).
\label{eq:hamiltonian}
\end{equation}
Here $\langle \bullet, \bullet \rangle$ denotes the scalar product and we call $\ell$ and $r$ the loss function and the regularizer of the problem respectively.

In this quite general context, the purpose of GASP is to approximately compute the marginal distribution $p(x_i)$, along with some expected quantities such as e.g. $\hat{\bx} = \mathbb{E}_p \bx$. 
The approximation entailed in GASP turns out to be exact under some assumptions in the large $N$ limit, as we shall later see.
A crucial assumption in the derivation of the GASP algorithm (and of GAMP as well), is that the entries of $\bF$ are independently generated according to some zero mean and finite variance distribution.   

Although the general formulation of GASP, presented in Sec. 2 of the SM, is able to deal with any model of the form \eqref{eq:p}, we will here restrict the setting to discuss Generalized Linear Estimation \cite{rangan2011generalized}.   

In GLE problems, $p(x)$ is sensibly chosen in order to infer a \emph{true signal} $\bx_0 \in \mathbb{R}^N$, whose components are assumed to be independently extracted from some prior $P_0$, $x_{0,i}\sim P_0\ \forall i\in [N]$.
The observations are independently produced  by  a (probabilistic) scalar channel $P^{\text{out}}$: $y_{\mu}\sim P^{\text{out}}(\bullet\,|\,\langle\bF^{\mu},\bx_{0}\rangle)$. 

It is then reasonable to choose $\ell\big(y,z) = -\log P^{\text{out}}(y|z)$,  $r(x) = -\log P_{0}(x)$ and $\beta=1$, so that the probability density $p(\bx)$ corresponds to the true posterior $P(\bx | \bF, \by)\propto P^{\text{out}}(\by| \bx, \bF)P_0(\bx)$, where $\propto$ denotes equality up to a normalization factor.
We refer to this setting as to the Bayesian-optimal or matched setting \cite{barbier2018optimal}.
Notice that in the  $\beta\uparrow\infty$ limit $p(x)$ concentrates around the maximum-a-posteriori (MAP) estimate. 
If $\beta\neq 1$ or if the Hamiltonian doesn't correspond to the minus log posterior (e.g, when $P_{0}$ and $P^{\text{out}}$ used in the Hamiltonian do not correspond to true ones) we talk about model mismatch. 

As a testing ground for GASP, and
the corresponding State Evolution, we here consider the phase retrieval problem, 
which has undergone intense investigation in recent years \cite{candes2015phase,dhifallah2017fundamental, chen2018gradient,goldstein2018phasemax,mondelli2018fundamental,sun2018geometric,mukherjee2018phase}.
We examine its noiseless and 
real-valued formulation, where observations are generated according to the process
\begin{align}
\bx_{0} &\sim\mathcal{N}(0,I_N),\\
F^{\mu}_i &\sim \mathcal{N}(0,1/N)\quad \forall \mu\in [M],\forall i\in [N], \\
y_{\mu} &\sim|\langle\bF^{\mu},\bx_{0}\rangle|.
\end{align}
for some $M$ and $N$, such that $\alpha=M/N >1$. 
For such generative model, we will focus on the problem of recovering
$\bx_0$ by minimizing the energy function $\mathcal{H}_{\by,\bF}(\bx)$ of Eq. \eqref{eq:hamiltonian}, in the case 
\begin{align}
 \ell(y,z) &=(y-|z|)^{2}, \label{eq:loss}\\   
 r(x) &=\frac{1}{2}\lambda\, x^{2}. \label{eq:regularizer}
\end{align}
Since the setting assumed for inference corresponds to MAP estimation in presence of a noisy channel, we are dealing with a case of model mismatch. The effect of the parameter $\lambda$ on the estimation shall be explored in Sec. \ref{sec:l2regularization},
 but we assume $\lambda=0$ until then. 
The optimization procedure will be performed using the zero-temperature (i.e. $\beta\uparrow\infty$) version of the GASP
algorithm.

\section{Previous work on Approximate Message Passing for Phase Retrieval}

Generalized approximate message passing (GAMP) was developed and rigorously
analyzed in Refs. \cite{rangan2011generalized} and \cite{javanmard2013state}. It was then applied for the first time to the (complex-valued) phase retrieval problem in Ref.~\cite{schniter2015compressive}. In Ref.~\cite{barbier2018optimal} the authors report an algorithmic threshold  for the perfect recovery of  $\alpha_{\text{alg}}\approx 1.13$, when using matched AMP on the real-valued version of the problem. This is to be compared to the information theoretic bound $\alpha_{\text{IT}}=1$.

The performance of GAMP in the MAP estimation setting, instead, was
investigated in Ref. \cite{ma2018approximate,ma2019optimization}.
A ``vanilla'' implementation of the zero temperature GAMP equations
for the absolute value channel was reported to achieve
perfect recovery for real-valued signals above $\alpha_{\text{alg}}\approx2.48$.
The authors were able to show that the algorithmic threshold
of GAMP in the mismatched case can however be drastically lowered by introducing
regularization a regularization term ultimately continued to zero. The AMP.A algorithm proposed in \cite{ma2018approximate,ma2019optimization} uses an adaptive $L_{2}$ regularization that improves the estimation threshold and also makes the algorithm more numerically robust  compensating a problematic divergence that appears in the message-passing equations (see Sec. 1.3 in the SM for further details).


Another important ingredient for AMP.A's performance is initialization:
in order to achieve perfect recovery one has to start from a configuration
that falls within the basin of attraction of the true signal, which
rapidly shrinks as the sampling ratio $\alpha$ decreases. A well-studied method
for obtaining a configuration correlated with the signal is \emph{spectral
initialization}, introduced and studied in Refs. \cite{jain2013low,candes2015phase,chen2015solving}:
in this case the starting condition is given by the principal eigenvector
of a matrix obtained from the data matrix $\bF$ and the
labels $\by$ passed through a nonlinear processing function.
The asymptotic performance of this method was analyzed in \cite{lu2017phase},
while the form of the optimal processing function was described in
\cite{mondelli2018fundamental,luo2019optimal}. However, since the SE description
is based on the assumption of the initial condition being uncorrelated
with the data, in AMP.A the authors revisited the method, proposing
a modification that guarantees ``enough independency'' while still
providing high overlap between the starting point and the signal.

With the combination of these two heuristics, AMP.A is able to reconstruct the signal down $\alpha_{\mathrm{alg}}\approx1.5$. In the present
paper we will show that, with a basic continuation scheme, the 1RSB
version of the zero temperature GAMP can reach the Bayes-optimal
threshold $\alpha_{\text{alg}}\approx1.13$ also in the mismatched
case, without the need of spectral initialization.

\subsection{GAMP equations at zero temperature}

Here we provide a brief summary of the  AMP
equations for the general graphical model of Eq. \eqref{eq:p}, in the $\beta\uparrow\infty$ limit. 
This is both to allow an easy comparison with our novel GASP algorithm and to introduce some notation that will be useful in the following discussion.
There is some degree of model dependence in the scaling of the messages when taking the zero-temperature limit: here we adopt the one appropriate for over-constrained models in continuous space. 
Details of the derivation  can be found in Sec. 1 of the SM, along with the specialization of the equations for phase retrieval. 

First, we introduce two free entropy functions associated to the input and
output channels ~\cite{rangan2011generalized}:
\begin{align}
\varphi^{\text{in}}(B,A)=  \max_{x}\ -r(x)-\frac{1}{2}Ax^{2}+B x \label{eq:min_in}\\
\varphi^{\text{out}}(\omega,V,y)=  \max_{u}\ -\frac{(u-\omega)^{2}}{2V}-\ell(y,u)\label{eq:min_out}.
\end{align} 
We define for convenience $\varphi^{\text{in},t}_i=\varphi^{\text{in}}(B_{i}^t,A^t)$ and $\varphi^{\text{out},t}_\mu =\varphi^{\text{out}}(\omega_{\mu}^t,V^{t-1},y_\mu)$. In our notation the GAMP message passing equations read:
\begin{align}
\omega^{t}_\mu & =\sum_{i}F_{i}^{\mu}\hat{x}^{t-1}_i-g^{t-1}_\mu\,V^{t-1}\\
g^t_\mu & = \partial_{\omega}\varphi^{\text{out},t}_{\mu}\\
\Gamma^t_\mu &= -\partial_{\omega}^{2}\varphi^{\text{out},t}_{\mu} \label{eq:singular}\\
A^t & = c _{\bF} \sum_\mu \Gamma_\mu^t\\
B^t_{i} & =\sum_{\mu}F_{i}^{\mu}g^t_\mu+\hat{x}^{t-1}_i\,A^t\\
\hat{x}^{t}_i & =\partial_{B}\varphi^{\text{in},t}_{i} \\ 
\Delta_i^t &= \partial_{B}^{2}\varphi^{\text{in},t}_{i} \\
V^t & =c _{\bF} \sum_i  \Delta_i^{t}
\end{align}
where $c_{\bF} = \frac{1}{M N}\sum_{\mu,i} (F^\mu_i)^2$.
It is clear from the equations that the two free entropy functions are supposed to be twice differentiable. This is not the case for phase retrieval, where GAMP encounters some non-trivial numerical stability issues: during the 
message-passing iterations one would have to approximately evaluate an 
empirical average of $\partial^2_\omega\varphi^{\text{out},t}_{\mu}$, containing Dirac's $\delta$-function. This is the problem encountered in AMP.A of Ref. \cite{ma2018approximate}. We will see that this problem is not present in GASP thanks to a Gaussian smoothing of the denoising function. 

\section{Generalized Approximate Survey Propagation}
The (G)ASP algorithm builds on decades of progress within the statistical physics community in understanding and dealing with  rough high-dimensional landscapes. The starting point for the derivation of the algorithm is the
partition function of $m$ replicas (or clones) of the system $\{ \bx^a\}_{a=1}^m$:
\begin{equation}
Z_{\by,\bF}^m = \int \prod_{a=1}^m \prod_{i=1}^N d x^a_i\ e^{-\beta \sum_{a=1}^m \mathcal{H}_{\by,\bF}(\bx^a)}.    \label{eq:replicated_measure}
\end{equation}
Note that, while this probability measure factorizes trivially, setting $m\neq1$ can introduce many important differences with respect to the standard case, both from the algorithmic and from the physics standpoints \cite{monasson1995structural,antenucci2019approximate}.

We write down the Belief Propagation (BP) equations associated to the replicated factor graph, where messages are probability distributions associated to each edge over the single-site replicated variables $\bar{\bx}_i=(x^1_i, \dots, x^m_i)$.  We make the assumption that the messages are symmetric under the group of replica indexes permutations.
This allows for a parametrization of the message passing that can be continued analytically to any real value of $m$. 
The resulting algorithm  goes under the name of 1RSB Cavity Method or, more loosely speaking, of Survey Propagation (with reference in particular to a zero temperature version of the 1RSB cavity method in discrete constraint satisfaction problems), and led to many algorithmic breakthroughs in combinatorial optimization on sparse graphical models \cite{mezard2002analytic,braunstein2005survey,krzakala2007gibbs}. 
One possible derivation of the (G)ASP algorithm is as the dense graph limit of the Survey Propagation equations, in the same way as AMP is obtained starting from BP. The derivation requires two steps. First, BP messages are projected by moment-matching  onto (replica-symmetric) multivariate Gaussian distributions on the replicated variables $\bar{\bx}\in \mathbb{R}^m$, which we express in the form
\begin{equation}
\nu(\bar{\bx})  \propto \int \dd h\ 
e^{-\frac{1}{2\Delta_{0}}(h-\hat{x})^{2}}\prod_{a=1}^m e^{-\frac{1}{2\Delta_{1}}(x^{a}-h)^{2}};
 \end{equation}
 Then, messages on the edges are conveniently expressed in term of single site quantities. 
 We note that,  some statistical independence assumptions on the entries of the measurement matrix are crucial for the derivation, as goes for AMP as well.
 While the starting point of the derivation assumed integer $m$, the resulting message passing can be analytically continued to any real $m$. Applying this procedure to the GLE graphical model of Eq. \eqref{eq:p} we obtain the GASP equations. 
Here we consider the  $\beta\uparrow \infty$ limit to deal with the MAP estimation problem. Details of the GASP derivation and the finite $\beta$ GASP equations are given in Sec. 2 of the SM.
Particular care has to be taken in the limit procedure, as a proper rescaling with $\beta$ is needed for each parameter. For instance, as the range of sensible choices for $m$ shrinks towards zero for increasing $\beta$, we rescale $m$ through the substitution $m \leftarrow m/\beta$. 

Relying on the definitions given Eqs.~\eqref{eq:min_in} and \eqref{eq:min_out}, we introduce the two 1RSB free entropies:
\begin{align}
\phi^{\text{in}}(B,A_{0},A_{1},m)= & \frac{1}{m}\log\int Dz\ e^{m\varphi^{\text{in}}(B+\sqrt{A_{0}}z,\,A_{1})}\label{eq:phi-in-2}\\
\phi^{\text{out}}(\omega,V_{0},V_1,y,m)= & \frac{1}{m}\log\int Dz\ e^{m\varphi^{\text{out}}(\omega+\sqrt{V_{0}}\,z,\,V_1,y)}\label{eq:phi-out-2}.
\end{align}
Here $\int D z$ denotes the standard Gaussian integration $\int dz\, \exp(-z^2/2)/\sqrt{2\pi}$. Using the shorthand notations $\phi^{\text{in},t}_{i}  =\phi^{\text{in}}(B^t_{i},A^t_{0},A^t_{1},m)$ and $\phi^{\text{out},t}_{\mu}= \phi^{\text{out}}(\omega_{\mu}^t,V^{t-1}_{0},V^{t-1}_1,y_{\mu},m)$ (notice the shift in the time indexes), and using again the definition $c_{\bF}=\frac{1}{M N}\sum_{\mu,i} (F^\mu_i)^2$ (hence $\mathbb{E} c_{\bF} = 1/N$ in our setting),  the GASP equations read:
\begin{align}
\omega^{t}_\mu & =\sum_{i}F_{i}^{\mu}\hat{x}^{t-1}_i-g^{t-1}_\mu\,(m V^{t-1}_{0}+V^{t-1}_{1})\label{eq:upd-w}\\
g^t_\mu & = \partial_{\omega}\phi^{\text{out},t}_{\mu}\label{eq:upd-g}\\[8pt]
\Gamma^{t}_{0,\mu} & =2\partial_{V_1}\phi^{\text{out},t}_{\mu}-(g^t_\mu)^{2}\label{eq:upd-gamma0}\\[8pt]
\Gamma^{t}_{1,\mu} & =-\partial_{\omega}^{2}\phi^{\text{out},t}_{\mu}+m\Gamma^{t}_{0,\mu}\label{eq:upd-gamma1}\\[10pt]
A^{t}_{0} & =c_{\bF} \sum_\mu \Gamma^{t}_{0,\mu}\label{eq:upd-A0}\\[-2pt]
A^{t}_{1} & =c_{\bF} \sum_\mu \Gamma^{t}_{1,\mu}\label{eq:upd-A1}\\
B^t_{i} & =\sum_{\mu}F_{i}^{\mu}g^t_\mu-\hat{x}^{t-1}_i\,(m A^{t}_{0}-A^{t}_{1})\label{eq:upd-B}\\
\hat{x}^{t}_i & =\partial_{B}\phi^{\text{in},t}_{i}\label{eq:upd-x}\\[8pt]
\Delta_{0,i}^{t} & =-2\partial_{A_{1}}\phi^{\text{in},t}_{i}-(\hat{x}^{t}_i)^{2}\label{eq:upd-delta0}\\
\Delta_{1,i}^{t} & =\partial_{B}^{2}\phi^{\text{in},t}_{i}-m\Delta_{0,i}^{t}.\label{eq:upd-delta1}\\
V^{t}_{0} & =c_{\bF}  \sum_i \Delta_{0,i}^{t} \label{eq:upd-V0}\\[-2pt]
V^{t}_{1} & =c_{\bF} \sum_i \Delta_{1,i}^{t} \label{eq:upd-V1}
\end{align}

\begin{algorithm}[tb]
   \caption{GASP($m$) for MAP} 
   \label{alg:gasp}
\begin{algorithmic}
   \STATE initialize $g_\mu = 0 \ \forall \mu$
   \STATE initialize $V_{0},\,V_{1},\, \hat{x}_i\ \forall i$ to some values 
   \FOR{$t=1$ {\bfseries to} $t_\text{max}$}
       \STATE compute $\omega_\mu, g_\mu, \Gamma^0_\mu,\Gamma^1_\mu  \; \forall \mu$ using (~\ref{eq:upd-w},\ref{eq:upd-g},~\ref{eq:upd-gamma0},~\ref{eq:upd-gamma1})  
	   \STATE compute $A_0,A_1$  using 
	   (\ref{eq:upd-A0},\ref{eq:upd-A1})
	   \STATE compute $B_i, \hat{x}_i ,\Delta_{0,i},\Delta_{1,i}\; \forall i$ using (~\ref{eq:upd-B},~\ref{eq:upd-x},~\ref{eq:upd-delta0},~\ref{eq:upd-delta1}) 
	   \STATE compute $V^0, V^1$ 
	   using (\ref{eq:upd-V0},~\ref{eq:upd-V1})
   \ENDFOR
\end{algorithmic}
\end{algorithm}

The computational time and memory complexity per iteration of the algorithm is the same of GAMP and is determined by the linear operations in Eqs. \eqref{eq:upd-w} and \eqref{eq:upd-B}. With respect to GAMP, we have the additional (but sub-leading) complexity due to the integrals in the input and output channels.
In some special cases, the integrals in Eqs.~\eqref{eq:phi-in-2} and \eqref{eq:phi-out-2} can be carried out analytically (e.g. in the phase retrieval problem).

Notice that GASP iteration reduces to standard GAMP iterations if $V_0$ and $A_0$ are initialized (or shrink) to zero, but can produce non-trivial fixed points depending on the initialization condition and on the value of $m$. 

We remark the importance of setting the time-indices correctly
in order to allow convergence \cite{caltagirone2014convergence}. The full algorithm is
detailed in Alg.~\ref{alg:gasp}.

The expressions for the special case of the absolute value channel \eqref{eq:loss}
and  $L_{2}$ regularization \eqref{eq:regularizer} can be found in Sec. 2.4 of the SM.
An important comment is that the divergence issue arising in AMP.A,
in the same setting, does not affect GASP: the discontinuity in the
expression for the minimizer of Eq.~\eqref{eq:min_out} is smoothed
out in the 1RSB version by the Gaussian integral in Eq.~\eqref{eq:phi-in-2}. We also note that, in phase retrieval, a problematic initialization
can be obtained by choosing configurations that are exactly orthogonal to the signal, 
since the message-passing will always be trapped in the uninformative 
fixed-point (due to the $Z_2$ symmetry of the problem). 
However, for finite size instances, a random Gaussian initial condition will have an overlap $\rho\equiv\langle \hat\bx, \bx_0\rangle$ of order $\text{\ensuremath{\mathcal{O}}}\left(1/\sqrt{N}\right)$ with the signal, which allows to escape the uninformative fixed point whenever it is unstable (i.e. for high $\alpha$).

\begin{figure}
\includegraphics[width=\columnwidth]{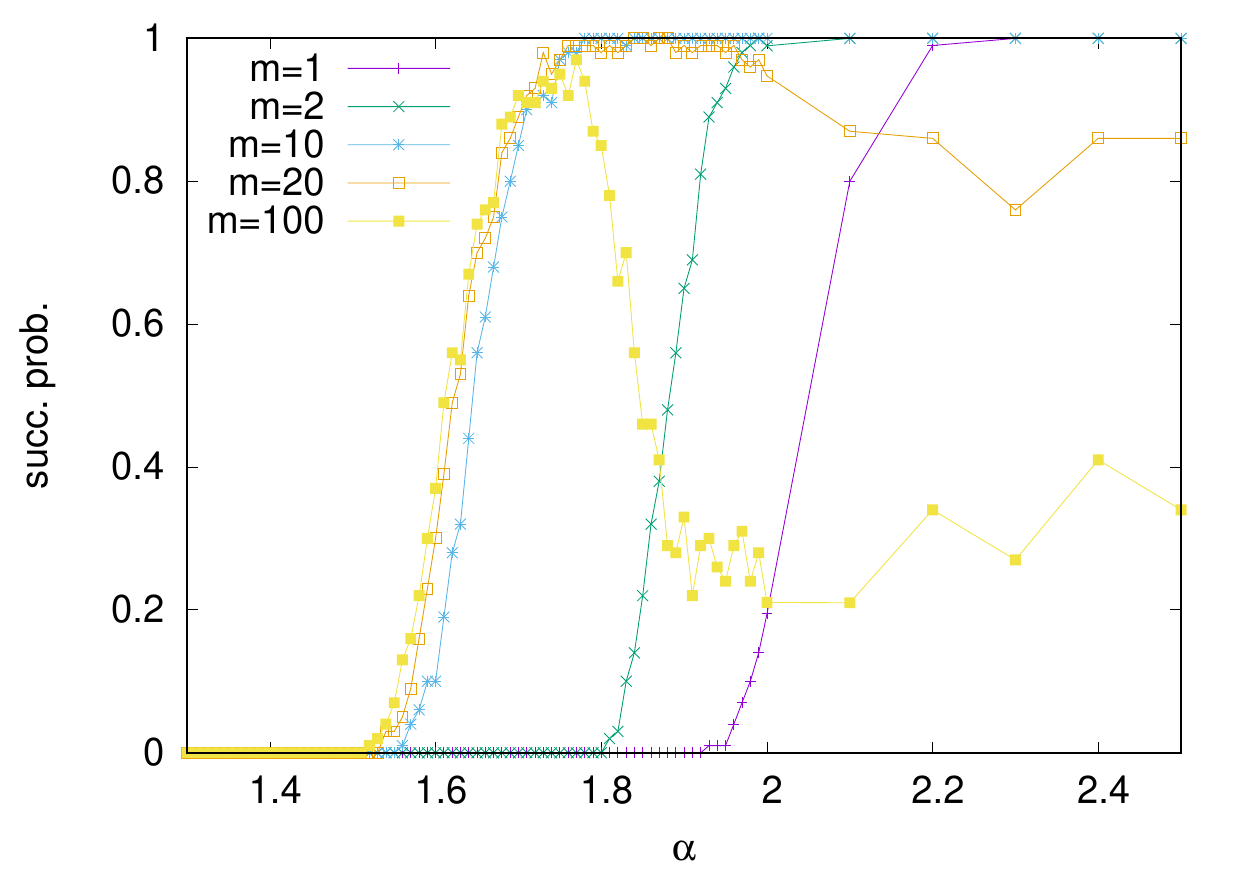}
\includegraphics[width=\columnwidth]{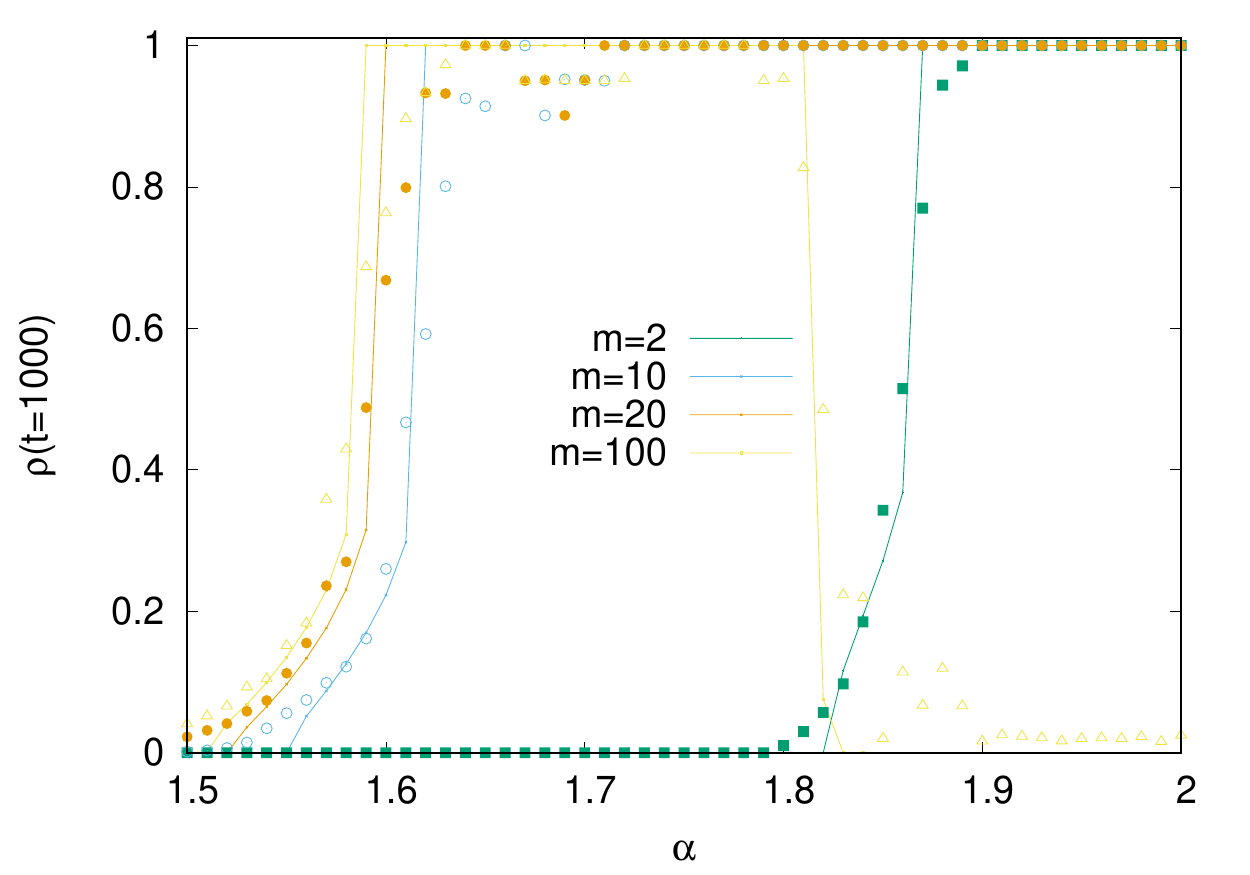}
\includegraphics[width=\columnwidth]{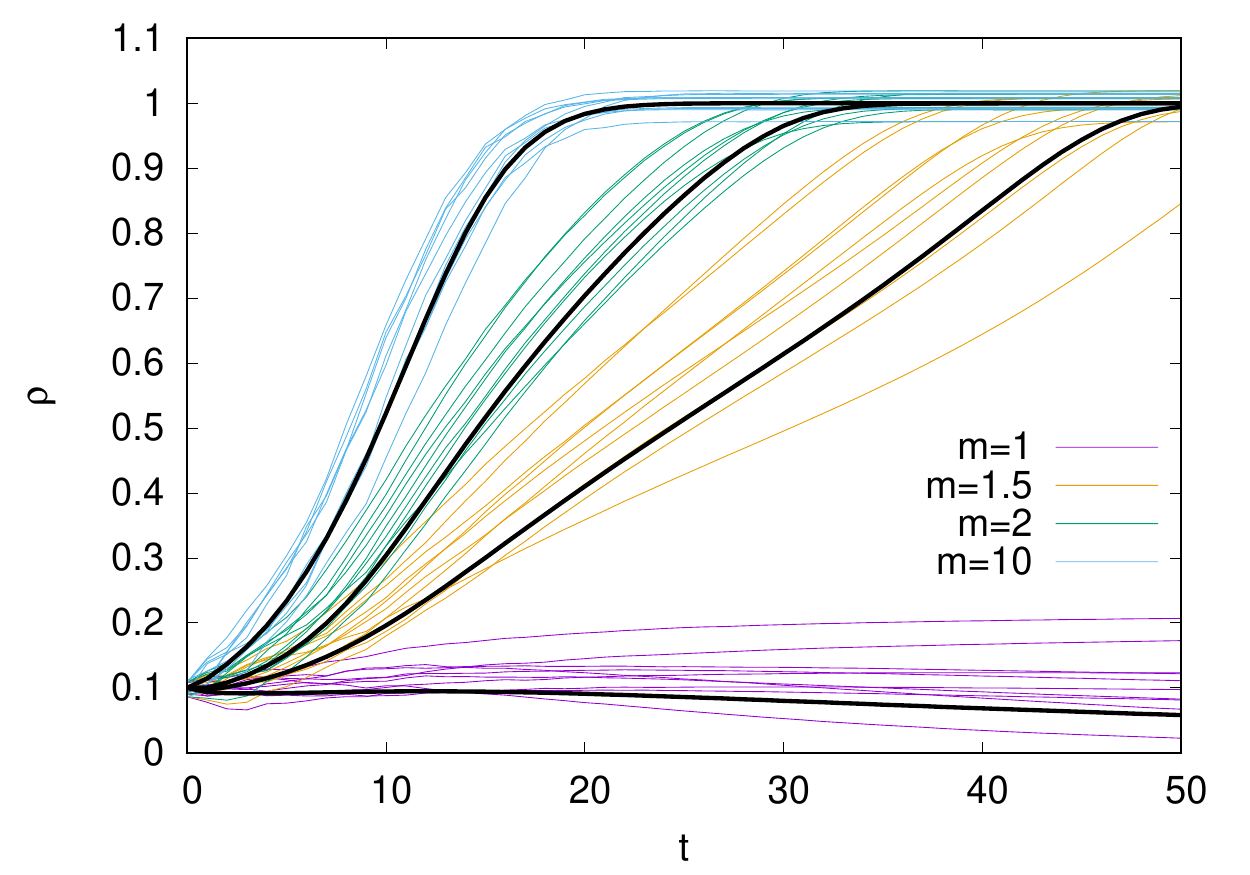}
\caption{\label{fig:gasp_se}(Top) Probability of perfect recovery of
the true signal using GASP (Alg. \ref{alg:gasp}), as a function of the sampling ratio $\alpha=M/N$.
(Middle) GASP and SE result after $t=10^3$ iterations. Start at $t=0$ with $\rho=10^{-3}$ for SE and $\hat{\bx}\sim\mathcal{N}(0,I_N)$ for GASP ($N=10^3$, averaged over $100$ samples).  (Bottom) Overlap $\rho^t$ with the true signal predicted  by SE dynamics at $\alpha=2$ and initial overlap $\rho=0.1$ (black lines) compared to 10 GASP trajectories for each value of $m$. Here  $N=10^4$.}
\end{figure}

In Fig.~\ref{fig:gasp_se} (Top and Middle),
we show the probability of a perfect recovery and convergence times
of GASP for the real-valued phase retrieval problem, for different sampling ratios $\alpha$ and values
of the symmetry-breaking parameter $m$, with $\lambda=0$. The initial condition is given by $V^{t=0}_0=V^{t=0}_1=1$ and $\hat{\bx}^{t=0}\sim\mathcal{N}(0,I_N)$. Notice that standard Gaussian initialization is able to break the symmetry of the channel and, at large $t$, GASP matches the fixed points predicted by SE (see next Section) with a small initial overlap with the true signal ($\rho^{t=0}=10^{-3}$). 
In order to achieve signal recovery at low $\alpha$, the symmetry-breaking
parameter has to be increased. In correspondence of values $m\approx100$, we 
report an algorithmic threshold around $\alpha_{\mathrm{alg}}^{\lambda=0}\approx1.5$. 
This threshold is comparable to the one of AMP.A, without exploiting adaptive regularization and spectral initialization as AMP.A (and which could be employed also for GASP). 

We report that, at fixed $m$, when $\alpha$ is increased above a certain value the message-passing will stop converging. The oscillating/diverging behavior of the messages can however be exploited for hand-tuning $m$, in the absence of a replica analysis to support the selection of its most appropriate value. More details can be found in Sec. 3 of the SM.

We presented here the zero-temperature limit of the GASP message-passing
to solve the MAP problem. Refer to Sec. 2 of the SM for a more general formulation dealing with the class of graphical models in the form of Eq. \ref{eq:p}. 

\section{State Evolution for GASP}
State Evolution (SE) is a set of iterative equations involving a few scalar quantities, that were rigorously proved to track the (G)AMP dynamics, in the sense of almost sure convergence of empirical averages \cite{javanmard2013state} in the large $N$ limit and with fixed sampling ratio $\alpha=M/N$. Following the analysis of Ref. \cite{rangan2011generalized} for GAMP, in order to present the SE equations for GASP  we assume that the observation model $y_{\mu}\sim P^{\text{out}}(\bullet\,|\,\langle\bF^{\mu},\bx_{0}\rangle)$
is such that can be expressed in the form 
$y^\mu \sim h(\langle \bF^\mu, \bx_0\rangle, \xi^\mu)$ for some function $h(z,\xi)$, with 
$\xi^\mu$ a scalar- or vector-valued random variable modeling the noise and sampled according to some distribution $P_\xi$. We also set $F^\mu_i\sim \mathcal{N}(0,1/N)$ i.i.d.. 
The recursion is a closed set of equations over the variables $\hat{\rho}^{t}, \hat{q}_{0}^{t}, A^{t}_0, A^{t}_1, \rho^{t}, q_{0}^{t}, V^t_0$ and $V^t_1$
Initializing at time $t=0$ the variables $\rho, q_0, V_0$ and $V_1$, the SE equations for $t\geq 1$: 
\begin{align} 
\hat{\rho}^{t}	&=\alpha\mathbb{E}\big[\partial_{z}\partial_{\omega}\phi^{\text{out}}(\omega^{t},V^{t-1}_0,V^{t-1}_1,h(z,\xi),m)\big]  \label{eq:se_rhoh}\\
\hat{q}_{0}^{t}	&=\alpha\mathbb{E}\big[\left(\partial_{\omega}\phi^{\text{out}}(\omega^{t},V^{t-1}_0,V^{t-1}_1,y,m)\right)^{2}\big]\\
A^{t}_0&=\alpha\mathbb{E}\big[2\partial_{V_1}\phi^{\text{out}}(\omega^{t},V^{t-1}_0,V^{t-1}_1,y,m)\big]-\hat{q}_{0}^{t} \\
A^{t}_1	&=\alpha\mathbb{E}\big[-\partial_{\omega}^{2}\phi^{\text{out}}(\omega^{t},V^{t-1}_0,V^{t-1}_1,y,m)\big]+m A^{t}_0,
\end{align}
where the expectation is over the process $\omega^t\sim	\mathcal{N}(0,q_{0}^{t-1})$, 
$z\sim	\mathcal{N}(\rho^{t-1}/q_{0}^{t-1}\,\omega^t,\,\mathbb{E}[x_{0}^{2}]-(\rho^{t-1})^{2}/q_{0}^{t-1})$, $\xi\sim P_\xi$ and $y \sim	h(z,\xi)$.
Also, we have a second set of equations that read:
\begin{align} 
\rho^{t}	&=\mathbb{E}\big[x_{0}\,\partial_{B}\phi^{\text{in}}(B^{t},A^{t}_0,A^{t}_1)\big] \\
q_{0}^{t}	&=\mathbb{E}\big[\big(\partial_{B}\phi^{\text{in}}(B^{t},A^{t}_0,A^{t}_1,m)\big)^{2}\big] \\
V^t_0	&=\mathbb{E}\big[-2\partial_{A_{1}}\phi^{\text{in}}(B^{t},A^{t}_0,A^{t}_1,m)\big]-q_{0}^{t} \\
V^t_1	&=\mathbb{E}\big[\partial_{B}^{2}\phi^{\text{in}}(B^{t},A^{t}_0,A^{t}_1)\big]-mV^t_0 
\label{eq:se_V1}
\end{align}
where the expectation is over the Markov chain $x_{0}\sim P_0$,
$B^{t} \sim\mathcal{N}(\hat{\rho}^{t}x_{0},\hat{q}_{0}^{t})$.

The trajectories of $V^t_0, V^t_1, A^t_0$ and $A^t_1$  in GASP concentrate  for large $N$ on their expected value given by the SE dynamics. In order to frame the GASP State Evolution in the rigorous setting of Ref.\cite{javanmard2013state}, we define a slightly different message-passing by replacing their GASP values for a given realization of the problem with the correspondent sample-independent SE values. Also, we replace $c_{\bF}$ with the expected value $1/N$. Let us define the denoising functions:
\begin{align}
\eta^{\text{out}}(\omega, y,t) &=
\partial_{\omega}\phi^{\text{out}}(\omega,V^{t-1}_0,V^{t-1}_1,y,m) \\
\eta^{\text{in}}(B, t) &=
\partial_{B}\phi^{\text{in}}(B,A^{t}_0,A^{t}_1,m)
\end{align}
and their vectorized extensions $\bet^{\text{out}}(\bw, \by,t) = (\eta^{\text{out}}(\omega_1, y_1,t),\dots, \eta^{\text{out}}(\omega_M, y_M,t))$ and $\bet^{\text{in}}(\bw, \by,t) = (\eta^{\text{in}}(B_1, t),\dots, \eta^{\text{in}}(B_N,t))$.
The modified GASP message-passing then reads
\begin{align}
\bw^t &= \bF \,\bet^{\text{in}}(\bB^{t-1}, t-1) - d^{\text{in}}_{t-1} \bet^{\text{out}}(\bw^{t-1},\by,t-1)  \label{eq:eta1}\\
\bB^t &= \bF^T \,\bet^{\text{out}}(\bw^{t},\by,t) - d^{\text{out}}_{t} \bet^{\text{in}}(\bB^{t-1}, t-1) \label{eq:eta2}
\end{align}
where the divergence terms are given by
\begin{equation}
\begin{aligned}
d^{\text{in}}_{t} &= \frac{1}{N}\sum_{i=1}^N \partial_{B} \eta^{\text{in}}(B_i^{t}, t)\\
d^{\text{out}}_{t} &= \frac{1}{N}\sum_{\mu=1}^M \partial_{\omega} \eta^{\text{out}}(\omega_\mu^{t},y_\mu,t)
\end{aligned}
\end{equation}
Message-passing (\ref{eq:eta1}, \ref{eq:eta2}) falls within the class of AMP algorithms analyzed in Ref. \cite{javanmard2013state} (under some further technical assumptions, see Proposition 5 there). Therefore, it can be rigorously tracked by the SE Eqs. (\ref{eq:se_rhoh},\ref{eq:se_V1}) in the sense specified in that work. In particular, denoting here $\hat\bx^t=\bet^{\text{in}}(\bB^{t}, t)$,  we have have almost sure converge in the large system limit of the overlap with the true signal and of the norm of $\hat\bx^t$ to their SE estimates:
 \begin{align}
\lim_{N\to\infty}\frac{1}{N}\langle \hat\bx^t, \bx_0\rangle &= \rho^t \\
\lim_{N\to\infty}\frac{1}{N} \langle \hat\bx^t, \hat \bx^t\rangle &= q_0^t
\end{align}
In Fig.~\ref{fig:gasp_se}(Bottom),
we compare the SE dynamics to the original GASP one (Alg. \ref{alg:gasp}). We compare SE
prediction for the evolution of the overlap $\rho$ to that 
observed in $10$ sample trajectories of GASP at $N=1000$, for a sampling ratio of $\alpha=2$ and different values of $m$. The initial estimate $\hat{\bx}^{t=0}$ in GASP was set to be a mixture $\hat{\bx}^{t=0}\sim\mathcal{N}(0,I_N)+0.1 \bx_0$. Therefore we initialize SE  with  $\rho^{t=0}=0.1$, and $q^{t=0}=1+(\rho^{t=0})^2$. Moreover, we set $V^{t=0}_0=V^{t=0}_1=1$ for both. As expected, we observe a good agreement between the two dynamics.  
 
\section{Effective Landscape and Message-Passing Algorithms}
The posterior distribution of statistical models in the hard phase is known to be riddled with glassy states \cite{antenucci2019glassy} preventing the retrieval of the true signal, a situation
which is exacerbated in the low temperature limit corresponding to MAP estimation. 

Within the replica formalism, the 1RSB free energy provides a description of this
complex landscape. The Parisi parameter $m$ allows to select the contributions of different families of 
states. More specifically $m$ acts as an inverse temperature coupled to the internal free energy of the states: increasing $m$ selects families of states with lower complexity (i.e., states that are less numerous) and lower free energy. 

The fixed points of the State Evolution of GASP are in one-to-one correspondence to the stationary points of the 1RSB free energy, and while the role of $m$ in the dynamics of SE is difficult to analyze, some insights can be gained from the static description given by the free energy. 

For phase retrieval in the MAP setting without regularization, a stable fixed-point of GAMP can be
found in the space orthogonal to the signal (i.e. at overlap $\rho=0$) for values of the sampling ratio below $\alpha^{\text{GAMP}} \approx 2.48$ \cite{ma2018approximate}, which is the algorithmic threshold for GAMP. 
For GASP instead, it is possible to see that the uninformative fixed-point is stable only below  
$\alpha^{\text{GASP}} \approx 1.5$, a noticeable improvement of the threshold with respect to GAMP.
This is obtained by choosing the $m$ corresponding to lowest complexity states according to the 1RSB free energy (see Sec. 3 of the SM for further details).
As we will see in the following, both these thresholds can be lowered by employing a continuation strategy
for the regularizer.

A thorough description of the results of the replica analysis and of the landscape 
properties for GLE models will be presented in a more technical future 
work.

\section{MAP estimation with an $L_{2}$ regularizer} \label{sec:l2regularization}

The objective function introduced in Eq.~\eqref{eq:hamiltonian} contains
a $L_2$ regularization term weighted by an intensity parameter $\lambda$.

Regularization plays and important role in reducing the variance of
the inferred estimator, and can be crucial when the observations are
noise-affected, since it lowers the sensitivity of the learned model
to deviations in the training set. However, as observed in \cite{ma2018approximate,ma2019optimization,balan2016reconstruction},
regularization is also useful for its smoothing effect, and can be
exploited in non-convex optimization problems even in the noiseless
setting. When the regularization term is turned up, the optimization
landscape gradually simplifies and it becomes easier to reach a global
optimizer. However, the problem of getting stuck in bad local minima
is avoided at the cost of introducing a bias.
The \emph{continuation} strategy is based on the fact that such biased 
estimator might be closer than the random initial configuration to the global optimizer of the unregularized objective : in a multi-stage approach, regularization is decreased (down to zero) after each warm restart.

Among the many possible continuation schedules for $\lambda$ (a little decrease after each minimization, or, as in AMP.A, at the end of each iteration) in this paper we choose a simple two-stage approach: first we run GASP till convergence with a given value of $\lambda>0$, then we set $\lambda=0$ in the successive iterations. 

\begin{figure}[ht]
\includegraphics[width=0.95\columnwidth]{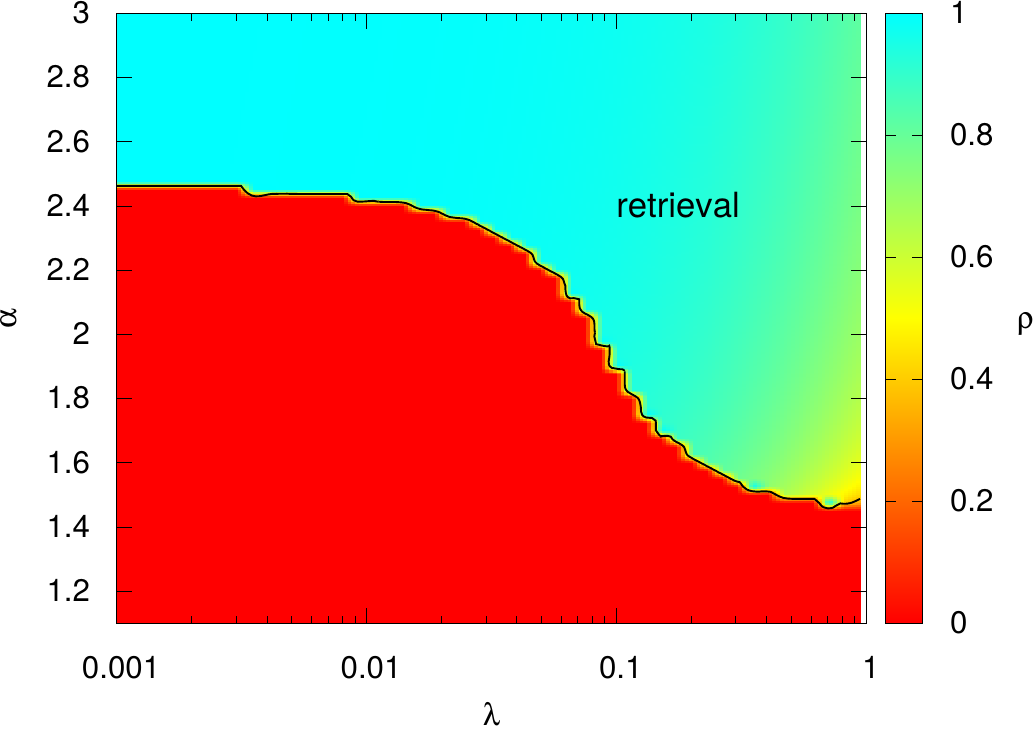}
\includegraphics[width=0.95\columnwidth]{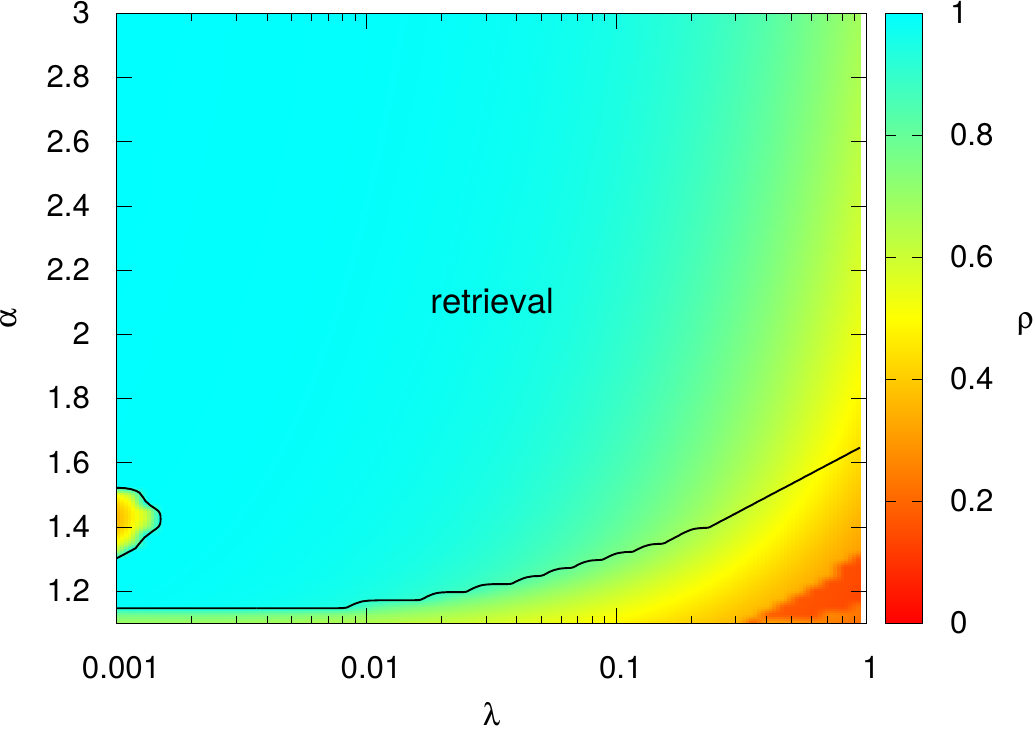}
\caption{\label{lambda_effect} Phase diagrams corresponding to the SE asymptotic 
analysis of GAMP (top) and GASP (bottom). The color maps indicate the 
overlap $\rho$ reached at convergence in the presence of an $L_2$ regularizer
of intensity $\lambda$.}
\end{figure}

In Fig.\ref{lambda_effect}, we compare the asymptotic performance 
(tracked by SE) of GAMP and GASP for the phase retrieval problem
with an $L_2$ regularization. The color map indicates the overlap with 
the signal reached at the end of the first stage of our continuation
strategy (with $\lambda\neq0$), while the black curves delimit the 
perfect retrieval regions, where the overlap reached at the end of 
stage two (with $\lambda=0$) is $\rho=1$. 

In both cases we set the initial variances $\Delta$ to $1$, and consider an 
initial condition with a small positive overlap with the signal, $\rho=0.1$.
An assumption of this kind is indeed needed to ensure that we avoid lingering on the 
fixed-point at $\rho=0$; however, the specific value of $\rho$ can be chosen 
arbitrarily (e.g., it could be taken much smaller without affecting
the phase diagrams). Even in real-world applications, it is often the case that
the non-orthogonality requirement is easily met, for example in many imaging
applications the signal is known to be real non-negative.
As explained in the previous section, we also set $q_{0}=1+\rho^{2}$ in the initialization of the self-overlap parameter.

In the GASP phase diagram, for each $\alpha$ and $\lambda$, the value of $m$ 
was set to the thermodynamic optimum value $m^{\star}$ (obtained at $\rho=0$), and was kept
fixed throughout the two stages of our continuation strategy. 
This $m^{\star}$ can be obtained by optimizing the 1RSB free energy 
over the symmetry-breaking parameter; the numerical values of $m$, 
corresponding to the points in the plot, can be found in Sec. 3 of the SM, in Fig. 1. It is not strictly necessary to fix $m$
to this specific value, as any value in a broad range of around $m^{\star}$ 
will still be effective (see for example Fig. 2 in the 
SM). As expected from the numerical 
experiments at $\lambda=0$, we can see from Fig.~\ref{lambda_effect} 
that when the regularizer becomes too small an uninformative fixed-point (in $\rho=0$) 
becomes attractive for the dynamics of GASP and signal recovery becomes impossible
below $\alpha_{alg}\sim1.5$ (we expect also the recovery region with $\alpha \in [1.13,1.3]$
at $\lambda=0.001$ to shrink and close when the regularizer is further decreased). 

It is clear that the introduction of an $L_2$-norm is crucial for 
reducing the algorithmic gap of both GAMP and GASP (the information theoretic 
threshold is $\alpha_{\mathrm{IT}}=1$), as previously observed
in \cite{ma2018approximate,ma2019optimization}. In this work we find that also in GLE problems, when the 
mismatched setting is considered (and inference happens off the 
Nishimori line \cite{nishimori2001statistical,antenucci2019approximate}), the more fitting
geometrical picture provided by the 1RSB ansatz can be exploited 
algorithmically: with a simple continuation strategy it is possible to lower the 
algorithmic threshold of GASP down to the Bayes-optimal value $\alpha=1.13$.

\section{Discussion}

We presented Generalized Approximate Survey Propagation, a novel algorithm designed to improve over AMP in the context of GLE inference problems, when faced with a mismatch between assumed and true generative model. The algorithm, parametrized by the symmetry-breaking parameter $m$, allows one to go beyond some symmetry assumptions at the heart of the previous algorithms, and proves to be more suited for the MAP estimation task considered in this work. 

In the prototypical case of real-valued phase retrieval, we have shown that with little tuning of $m$ it is possible to modify the effective landscape explored during the message-passing dynamics and avoid getting stuck in otherwise attractive uninformative fixed points. 
Furthermore, we have seen that, even in the noiseless case, a simple continuation strategy, based on the introduction of an $L_2$ regularizer, can guide GASP close enough to the signal and allow its recovery, extending the region of parameters where GASP is more effective than GAMP. In some cases we observed that GASP can achieve perfect retrieval until the Bayes-optimal threshold, at the sampling ratio $\alpha\sim1.13$. 
We also derived the 1RSB State Evolution equations, and showed that they can be used as a simple tool for tracking the asymptotic behaviour of GASP.

We delay a comprehensive analysis of the landscape associated to GLE models to a more technical publication, where we will also deal with the case of noisy observation channels. 
A straightforward follow-up of the present work could focus on the search for an adaptation scheme for the $L_2$ regularizer, possibly extending the work of Refs. \cite{ma2018approximate,ma2019optimization}, and more importantly, for a criterion to identify the best setting for the symmetry-breaking parameter.
Another possible future line of work could go in the direction of relaxing some of the assumptions made in deriving the GASP algorithm over the observation matrix. This could motivate the derivation of a 1RSB version of the Vector Approximate Message Passing equations \cite{schniter2016vector}.
Also, the extension of GASP to deep non-linear inference model, along the lines of Ref. \cite{manoel2017multi,fletcher2018inference} seems to be promising and technically feasible.

CL thanks Junjie Ma for sharing  and explaining the code of their AMP.A algorithm.

\bibliography{bibliography}
\bibliographystyle{icml2019}

\appendix
\onecolumn
\title{Supplementary Material}
\maketitle
\section{A recap on Generalized Approximate Message Passing \label{app:GAMP}}
\subsection{Derivation of GAMP \label{app:GAMPderiv}}

For the reader's convenience and for familiarizing with the notation adopted throughout this work, we sketch the derivation of the Generalized Approximate Message Passing (GAMP) equations
for Generalized Linear Estimation (GLE) models. For a longer discussion, we refer the reader to Refs. 
\cite{rangan2011generalized,ma2018approximate,kabashima2016phase}.
We assume the setting of Eq. (1) of the Main Text, that is a graphical model defined by the Hamiltonian:
\begin{equation}
\mathcal{H}_{\by,\bF}(\bx)=\sum_{\mu}\ell\big(y_{\mu},\, \langle \bF^{\mu}, \bx\rangle \big)+\sum_{i}r(x_{i}), \label{eq:model}
\end{equation}
with the further assumption that the entries of $\bF$ are i.i.d. zero-mean Gaussian variables with variance $1/N$, i.e  $F^{\mu}_i\sim\mathcal{N}(0,1/N)$ (but the derivation also applies to non-Gaussian variables with the same mean and variance). The configuration space is assumed to be some subset $\chi^N$ of $\mathbb{R}$. For discrete spaces, integrals should be replace with summations. 
Also, we consider the regime of large $M$ and $N$, with finite $\alpha=M/N$.
The starting point for the derivation of GAMP equations is the Belief Propagation (BP) algorithm \cite{mezard2009information}, characterized by the exchange of two sets of  messages: 
\begin{align}
\nu^{t}_{i\to\mu}(x_{i}) & \propto e^{-\beta r(x_{i})+\sum_{\nu\neq\mu}\log\hat{\nu}^{t}_{\nu\to i}(x_{i})} \label{eq:nu}\\
\hat{\nu}^{t+1}_{\mu\to i}(x_{i}) & \propto\int_{\chi^{N-1}}\prod_{j\neq i}d\nu^t_{j\to\mu}(x_{j})\ e^{-\beta \ell(y_{\mu},\langle \bF^{\mu},\bx\rangle)} . \label{eq:nu_hat}
\end{align}
For the dense graphical model we are considering, by virtue of central limit arguments, we can
relax the resulting identities among probability densities to relations among their first and second moments.
The resulting approximated version of BP goes under the name of relaxed Belief Propagation (rBP)
 \cite{guo2006asymptotic, rangan2010estimation, mezard2017mean}.

We define the expectations over the measure in Eq.\eqref{eq:nu} as $\left<\bullet \right>^t_{i\to\mu}$, and its moments as $\left<x\right>^t_{i\to\mu}=\hat{x}^t_{i\to\mu}$ and $\left<x^2\right>^t_{i\to\mu}=\Delta^t_{i\to\mu} + (\hat{x}^t_{i\to\mu})^2$. In high dimensions we can see that the scalar product $\langle \bF^{\mu}, \bx\rangle $ in Eq.\eqref{eq:nu_hat} becomes Gaussian distributed according to $\mathcal{N}(\sum_j F^{\mu}_j \hat{x}^t_{j\to\mu}+F^{\mu}_i(x_i-\hat{x}^t_{i\to\mu}), \sum_{j\neq i} (F^\mu_j)^2 \Delta^t_{j\to\mu})$. 

In order to obtain the relationship between the moments of the two sets of distributions it is useful to introduce two scalar estimation functions, the input and output channels, that fully characterize the problem. The associated free entropies \cite{barbier2018optimal} (i.e., log-normalization factors) can be expressed as:
\begin{align}
\varphi^{\text{in}}(B,A)= & \log \int_{\chi} dx\ e^{-\frac{1}{2}A x^{2}+ B x -\beta \ r(x)} \\
\varphi^{\text{out}}(\omega, V, y) = & \log \int\frac{dz}{\sqrt{2\pi V}}\ e^{-\frac{1}{2V}(z-\omega)^{2}-\beta\ \ell(y,z)}.
\end{align}
Then, defining $g^t_{\mu}=\partial_{\omega} \varphi^{\text{out}}(\omega',V',y)$ and $\Gamma^t_{\mu}=-\partial^2_{\omega} \varphi^{\text{out}}(\omega',V',y)$, both evaluated in $\omega'=\sum_{j} (F^\mu_j)^2 \Delta^t_{j\to\mu}$ and $V'=\sum_j F^{\mu}_j \hat{x}^t_{j\to\mu}$, we can express through them the approximate message-passing, obtained at the second order of the Taylor expansion of the  messages:
\begin{align}
\log\hat{\nu}^{t+1}_{\mu\to i}(x_{i}) = & \varphi^{\text{out}}\left(\sum_j F^{\mu}_j \hat{x}^t_{j\to\mu}+F^{\mu}_i(x_i-\hat{x}^t_{i\to\mu}),\ \sum_{j\neq i} (F^\mu_j)^2 \Delta^t_{j\to\mu}, \ y_\mu\right) + \text{const}. 
\end{align}

Next, we  close the equations on single site quantities, discarding terms which are sub-leading for large $N$ and assuming zero mean and 1/N variance i.i.d entries in $\bF$ . Thus, we can remove the cavities and approximate the parameters of the (non-cavity) estimation channels as follows:
\begin{align}
B^t_i = & \sum_{\mu}F^{\mu}_i g_{\mu}^t - \hat{x}^{t-1}_{i} \sum_{\mu}(F^{\mu}_i)^2 \Gamma^{t}_{\mu}  \\
A^t_i = & \sum_{\mu}(F^{\mu}_i)^2 \Gamma^t_{\mu} \\
\omega^t_{\mu} = &  \sum_i F^{\mu}_i \hat{x}^t_{i} -g^t_\mu \sum_i (F^{\mu}_i)^2  \Delta^t_i \\
V_\mu^t = & \sum_{i} (F^\mu_i)^2 \Delta_{i}.
\end{align}
Finally, the expectations introduced above can be obtained via the derivatives:
\begin{align}
g_\mu^t & = \partial_{\omega}\varphi^{\text{out},t}_{\mu}\\
\Gamma_\mu^t &= -\partial_{\omega}^{2}\varphi^{\text{out},t}_{\mu} \\
\hat{x}_i^t & =\partial_{B}\varphi^{\text{in},t}_{i} \\ 
\Delta_i^t &= \partial_{B}^{2}\varphi^{\text{in},t}_{i},
\end{align}
where we used the shorthand notation $\varphi^{\text{in},t}_i=\varphi^{\text{in}}(B_{i}^t,A^t)$ and $\varphi^{\text{out},t}_\mu =\varphi^{\text{out}}(\omega_{\mu}^t,V^{t-1},y)$.

A slight simplification of the message passing (which involves   $\mathcal{O}(N^2)$ operations per iteration), relies on the observation that due to the statistical   properties of $\bF$ the quantities $A_i$ and $V_\mu$ do not depend on their indexes \cite{rangan2011generalized}, so we can define their scalar counterparts:
\begin{align}
A^t & = c _{\bF} \sum_\mu \Gamma_\mu^t,\\
V^t & =c _{\bF} \sum_i  \Delta_i^{t-1},
\end{align}
where $c_{\bF} = \sum_{\mu,i} (F^\mu_i)^2 / (M N)\approx 1/N$. 
Therefore we obtain:
\begin{align}
\omega^{t}_\mu & =\sum_{i}F_{i}^{\mu}\hat{x}^{t-1}_i-g^{t-1}_\mu\,V^{t-1}\label{eq:gamp_w}\\
g^t_\mu & = \partial_{\omega}\varphi^{\text{out},t}_{\mu} \label{eq:g_scalar}\\
\Gamma^t_\mu &= -\partial_{\omega}^{2}\varphi^{\text{out},t}_{\mu} \label{eq:Gamma_scalar}\\
A^t & = c _{\bF} \sum_\mu \Gamma_\mu^t\\
B^t_{i} & =\sum_{\mu}F_{i}^{\mu}g^t_\mu+\hat{x}^{t-1}_i\,A^t\\
\hat{x}^{t}_i & =\partial_{B}\varphi^{\text{in},t}_{i} \label{eq:xhat_scalar} \\ 
\Delta_i^t &= \partial_{B}^{2}\varphi^{\text{in},t}_{i} \label{eq:Delta_scalar}  \\
V^t & =c _{\bF} \sum_i  \Delta_i^{t}.\label{eq:gamp_V}
\end{align}
Eqs. (\ref{eq:gamp_w}-\ref{eq:gamp_V}) are known as the GAMP iterations, and are valid for $t\geq 1$, given some initial condition $\hat{\bx}^{t=0}$ and $V^{t=0}$, along with $g^{t=0}_\mu = 0, \ \forall \mu$.

\subsection{Zero-temperature limit of GAMP}
In order to apply the GAMP algorithm to MAP estimation or MAP + regularizer, we have to consider the zero-temperature limit $\beta\uparrow\infty$ . The limiting form of the equations depends on the model and on the regime (e.g. low or high $\alpha$). Here we consider models defined on continuous spaces $\chi^N$ and  in the high $\alpha$ regime (e.g. $\alpha >1$ for phase retrieval). In this case, while taking the limit, the message have to be rescaled appropriately in order for them  to stay finite. Therefore we rescale the messages through the substitutions:
\begin{align}
A & \to\beta A\\
B_i & \to\beta B_i\\
V & \to V/\beta\\
g_\mu & \to\beta g_\mu\\
\Delta_i & \to\Delta_i/\beta.
\end{align}
With these rescalings, the GAMP equations (\ref{eq:gamp_w}-\ref{eq:gamp_V}) are left unaltered, but the expressions for the free entropies of the scalar channels become
\begin{align}
\varphi^{\text{in}}(B, A)= & \max_{x\in \chi}\ -r(x)-\frac{1}{2}A x^{2} + B x \label{eq:varphi_in_Binf} \\
\varphi^{\text{out}}(\omega, V ,y) = &\max_{z}\ -\frac{(z-\omega_\mu)^{2}}{2V}-\ell(y,z), \label{eq:varphi_out_Binf}
\end{align}
as it is easy to verify.

\subsection{GAMP equations for real-valued phase retrieval and AMP.A equations}
In the special case of the phase retrieval problem, with a loss $\ell(y,\omega) = (y-|\omega|)^2$ and $L_2$-norm $r(x)=\lambda x^2 / 2$ and at zero temperature, the two scalar estimation
channels of Eqs.\eqref{eq:varphi_in_Binf} and \eqref{eq:varphi_out_Binf} become:
\begin{align}
\varphi^{\text{in}}(B,A)= & \frac{B^2}{2(A+\lambda)} \label{eq:varphi_in_PR}  \\
\varphi^{\text{out}}(\omega,V,y)= & -\frac{(y-|\omega|)^2}{2V+1}. \label{eq:varphi_out_PR}
\end{align}
Thus, Eqs. (\ref{eq:g_scalar}, \ref{eq:xhat_scalar}, \ref{eq:Delta_scalar}, \ref{eq:gamp_V}) simply yield:
\begin{align}
g^t_\mu & = \frac{2(y_\mu - |\omega_\mu^t|)}{2V^t+1}\sign(\omega_\mu^t) \\
\hat{x}^{t}_i & = \frac{B^t_i}{A^t+\lambda} \\ 
\Delta_i^t &= \frac{1}{A^t+\lambda} \\
V^t &= N c_{\bF} \frac{1}{A^t+\lambda} . 
\end{align}
Eq. \eqref{eq:Gamma_scalar} is instead singular, since it involves the derivative of the $\sign$ function. Since we have
\begin{align}
\omega^{t}_\mu & =\sum_{i}F_{i}^{\mu}\hat{x}^{t-1}_i-\,\frac{g^{t-1}_\mu}{A^{t-1}+\lambda}\\
g^t_\mu & = \frac{2(y_\mu - |\omega_\mu^t|)}{2V^t+1}\sign(\omega_\mu^t)\\
A^t & = -c_{\bF} \sum_\mu \partial_{\omega}^2\varphi_\mu^{{\text{out}},t} \\
x^t_{i} & = (A^t+\lambda) \left(\sum_{\mu}F_{i}^{\mu}g^t_\mu+\hat{x}^{t-1}_i\,A^t\right),
\end{align}
because of the singularity, the value of $A^t$ cannot be simply evaluated on a given finite sample. A possible way of dealing with this issue is to use a smoothing strategy in the first iterations of the message passing, replacing the $\sign$ function with a continuous version of it. Alternatively, in Ref. \cite{ma2018approximate,ma2019optimization}, the author propose to self-consistently adapt the regularizer $\lambda$ at each time step in order to absorb the divergent contribution. Also, the dynamics $A^t$  can be replaced by the corresponding and non-singular SE estimate.
We find that all these solutions are difficult to implement in a robust way and lead to some numerical instabilities that have to be dealt with great care. As we commented in the Main Text, this problem does not affect the GASP version of the algorithm, because of the  additional Gaussian kernel that smoothens the output scalar estimation channel.  

\section{Derivation of Generalized Approximate Survey Propagation \label{app:GASP}}

We will derive the GASP equation for a general GLE model specified by \eqref{eq:model}. As already explained in the Main Text, we will follow \cite{antenucci2019approximate} and work within the (real) replicas formalism. 
The derivation is similar to the one outlined for the GAMP algorithm, which goes from Belief Propagation (BP) to relaxed Belief Propagation (rBP) to Approximate Message Passing (AMP). In fact, GASP is obtained by applying the very same procedure that leads to GAMP to an auxiliary graphical model that corresponds to considering multiple copies of the system. 

\subsection{Relaxed Survey Propagation}
As an intermediate step toward the derivation of GASP equations, we derive the relaxed Survey Propagation (rSP) equations for out GLE problem. This corresponds to a Gaussian closure of the standard BP equations on the replicated factor graph of the problem, under replica symmetric assumptions.
We assume the setting of Eq. (1) of the Main Text, that is a graphical model defined by the Hamiltonian:
\begin{equation}
\mathcal{H}_{\by,\bF}(\bx)=\sum_{\mu}\ell\big(y_{\mu},\, \langle \bF^{\mu}, \bx\rangle \big)+\sum_{i}r(x_{i}), 
\end{equation}
with the further assumption that the entries of $\bF$ are i.i.d. zero-mean Gaussian variables with variance $1/N$, i.e  $F^{\mu}_i\sim\mathcal{N}(0,1/N)$ (but the derivation also applies to non-Gaussian variables with the same mean and variance). The configuration space is assumed to be some subset $\chi^N$ of $\mathbb{R}$. For discrete spaces, integrals should be replaced with summations. 
Also, we consider the regime of large $M$ and $N$, with finite $\alpha=M/N$.

Quite peculiarly, the family of message passing algorithm corresponding to the 1RSB framework (i.e. SP, rSP, ASP), are simply obtained as the BP, rBP and AMP equations for a \emph{replicated} graphical model,
\begin{equation}
p(\{\bx^a\}_{a=1}^m) = \frac{1}{Z^m_{\by,\bF} }e^{-\beta\sum_{a=1}^m \mathcal{H}_{\by,\bF}(\bx^a)}, \label{eq:p_rep}
\end{equation}
where $m$ is the number of  replicas. The parameter $m$ is not to be confused with the number of replica $n$ that it is usually sent to zero in the replica trick, but it has to be interpreted as the  Parisi symmetry breaking parameter  in the 1RSB scheme or as the number of real clones within Monasson's method \cite{monasson1995structural}).
While the replicated model is trivially factorized over the replicas, a highly non-trivial picture emerges when $p$ is considered as the limit distribution obtained by inserting a coupling term among the replicas and then letting it go to zero. Since the discussion about this technique (pioneered by Monasson in Ref. \cite{monasson1995structural}) is quite articulated and has its root in a few decades of development in spin-glass theory, we refer the interested reader to  \cite{mezard1987spin,mezard2009information,antenucci2019glassy,antenucci2019approximate} and reference therein for an overview of the theoretical aspects behind this approach. From here on we present the innovative aspects of our contribution, which extends the work of  Ref. \cite{antenucci2019approximate} to GLE models.

We denote with $\bx_i\in \xi^m$ the replicated variable on site $i$, and write a first set of BP equations in the form:
\begin{align}
\nu_{i\to\mu}(\bar{\bx}_{i}) & \propto e^{-\beta\sum_{a=1}^m r(x_{i}^{a})+\sum_{\nu\neq \mu}\log\hat{\nu}_{\nu\to i}(\bar{\bx}_{i})}, \label{eq:gasp_nu}
\end{align}
where we omit time indexes.
In the large $N$ limit, we can exploit the statistical assumptions on $\bF$ and the central limit theorem to perform a Gaussian approximation of the messages. Also, we assume symmetry of the messages $\nu_{i\to\mu}(\bar{\bx}_{i})$ under permutation of replica indexes,  which holds self-consistently if one makes a similar assumption also on the messages $\hat{\nu}_{\nu\to i}(\bar{\bx}_{i})$. Messages are then multivariate Gaussian distribution conveniently parametrized by the mean $\hat{x}_{i\to\mu}$ and two parameters $\Delta_{0, i\to\mu}$ and $\Delta_{1, i\to\mu}$ in the form:
\begin{align}
 \nu_{i\to\mu}(\bar{\bx}_{i}) &  \propto\int \dd h\ e^{-\frac{1}{2\Delta_{0, i\to\mu}}(h-\hat{x}_{i\to\mu})^{2}}\prod_{a}e^{-\frac{1}{2\Delta_{1, i\to\mu}}(x_{i}^{a}-h)^{2}},
 \end{align}
also known as caging ansatz in the glass and spin-glass community \cite{charbonneau2017glass}.
According to this Gaussian projection, the first and second moments of messages are given by
\begin{align}
\langle x_{i}^{a}\rangle_{i\to\mu} & =\hat{x}_{i\to\mu}\\
\langle x_{i}^{a}x_{i}^{b}\rangle_{i\to\mu} & =\Delta_{0, i\to\mu}+\hat{x}_{i\to\mu}^{2}\\
\langle(x_{i}^{a})^{2}\rangle_{i\to\mu} & =\Delta_{1, i\to\mu}+\Delta_{0, i\to\mu}+\hat{x}_{i\to\mu}^{2}.
\end{align}

The values of $\hat{x}_{i\to\mu}$, $\Delta_{0, i\to\mu}$ and $\Delta_{1, i\to\mu}$ can be obtained by matching the moments of the r.h.s. of \ref{eq:gasp_nu}. 
From now on the derivation is very close to that  of Section \ref{app:GAMPderiv} for GAMP, therefor we relax the notation and drop some indexes. 
Let us define the input channel free entropy:
\begin{equation}
\phi^{\text{in}}(B, A_{0},A_{1},m)=\frac{1}{m}\log\int Dz\,\left(\int_\chi dx\ e^{-\beta r(x)-\frac{1}{2}A_{1}x^{2}+(B+\sqrt{A_{0}}z)x}\right)^{m}. \label{eq:phi_in_1RSB}
\end{equation}

Let us also denote with $\langle \psi(\bar{\bx}) \rangle$ the expectation over the corresponding measure, in the $m$-replicated space, of a test function $\psi$, that is
\begin{equation}
\langle \psi(\bar{\bx}) \rangle = \frac{\int Dz\int_{\chi^{m}}\prod_{a=1}^m dx^a\ e^{-\beta r(x^a)-\frac{1}{2}A_{1}(x^a)^{2}+(B+\sqrt{A_{0}}z)x^a}\ \psi(\bar{\bx})}
{\int Dz\prod_{a=1}^m dx^a\ e^{-\beta r(x^a)-\frac{1}{2}A_{1}(x^a)^{2}+(B+\sqrt{A_{0}}z)x^a}}.
\end{equation}
For appropriate values of $B_{\nu\to i}, A_{0, \nu\to i}$ and $A_{1, \nu\to i}$ to be determined by second order expansion of  $\log\hat{\nu}_{\nu\to i}(\bar{x}_{i})$, and for replica indexes $a$ and $b$,$a\neq b$, from Eq. \eqref{eq:gasp_nu}  we obtain:
\begin{align}
\partial_{B}\phi^{\text{in}} & =\langle x^{a}\rangle\\
\partial_{B}^{2}\phi^{\text{in}} & =(\langle(x^{a})^{2}\rangle-\langle x^{a}x^{b}\rangle)+m\left(\langle x^{a}x^{b}\rangle-\langle x^{a}\rangle^{2}\right)\\
2\partial_{A_{0}}\phi^{\text{in}} & =(\langle(x^{a})^{2}\rangle-\langle x^{a}x^{b}\rangle)+m\langle x^{a}x^{b}\rangle\\
2\partial_{A_{1}}\phi^{\text{in}} & =-\langle(x^{a})^{2}\rangle.
\end{align}

Using the above formulas, we can project the measure on $\mathbb{R}^m$ corresponding to $\phi^{\text{in}}$ onto the space of replica-symmetric Gaussian distributions, parametrized  by $\hat{x}$, $\Delta_0$ and $\Delta_1$. Defining for convenience $\phi^{\text{in}}_{i\to\mu}=\phi^{\text{in}}(A_{0,i\to \mu},A_{1,i\to \mu},B_{i\to \mu})$, with the quantities $A_{0,i\to \mu},A_{1,i\to \mu}$ and $B_{i\to \mu}$ to be defined later, by moment matching we obtain:
\begin{align}
\hat{x}_{i\to \mu} & = \partial_{B}\phi^{\text{in}}_{i\to\mu}, \label{eq:rSP-x}\\
\Delta_{0,i\to\mu} & =\frac{1}{m-1}\left(\partial_{B}^{2}\phi^{\text{in}}_{i\to\mu}+2\partial_{A_{1}}\phi^{\text{in}}_{i\to\mu}+\hat{x}_{i\to\mu}^2\right),\\
\Delta_{1,i\to\mu} & =\partial_{B}^{2}\phi^{\text{in}}_{i\to\mu}-m \Delta_{0,i\to\mu}.
\end{align}
Defining the messages
\begin{align}
\omega_{\mu\to i} & =\sum_{j\neq i}F_{j}^{\mu}\hat{x}_{j\to\mu},\\
V_{0, \mu\to i} & =\sum_{j\neq i}\left(F_{j}^{\mu}\right)^{2}\Delta_{0, j\to\mu},\\
V_{1, \mu\to i} & =\sum_{j\neq i}\left(F_{j}^{\mu}\right)^{2}\Delta_{1, j\to\mu}, \label{eq:rSP-V1}
\end{align}
we can express the central limit approximation for the BP equations at factor node $\mu$ as
\begin{align}
\hat{\nu}_{\mu\to i}(\bar{\bx}_{i}) & \propto\int_{\chi^{m(N-1)}}\prod_{j\neq i}\dd\nu_{j\to\mu}(\bar{\bx}_{j})\ e^{-\beta\sum_{a}\ell(y_{\mu},\langle \bF^{\mu}, \bx^{a}\rangle)}\\
 & \propto\int \dd z_{0}\ e^{-\frac{1}{2V_{0, \mu\to i}}(z_{0}-\omega_{\mu})^{2}}\prod_{a=1}^m\left(\int Dz_{1}\ e^{-\beta\ \ell(y_{\mu},F_{i}^{\mu} x_{i}^{a}+z_{0}+\sqrt{V_1}z_{1})}\right).
\end{align}
The expansion of the  message $\hat{\nu}_{\mu\to i}(\bar{\bx}_{i})$ that we use for our Gaussian closure of the BP messages are conveniently expressed in terms of the derivatives of the output channel free entropy
\begin{align}
\phi^{\text{out}}(\omega, V_0, V_1, y,m) & =\frac{1}{m}\log\int\frac{\dd z_0}{\sqrt{2\pi V_{0}}}\ e^{-\frac{1}{2V_{0}}(z_0-\omega)^{2}}\ \left(\int Dz_{1}\ e^{-\beta\ \ell(y,z_0+\sqrt{V_1}z_{1})}\right)^{m}. \label{eq:phi_out_1RSB}
\end{align}
Introducing the second order expansion
\begin{equation}
\log\hat{\nu}_{\mu\to i}(\bar{\bx}_{i}) = g_{\mu\to i} \sum_a x_i^a -\frac{1}{2} A_{1,{\mu\to i}} \sum_a (x^a_i)^2 +\frac{1}{2} A_{0,{\mu\to i}} \sum_{a,b} x^a_i x^b_i
\end{equation}
we can write the last set of rSP messages as 
\begin{align}
g_{\mu\to i} & = \partial_{\omega}\phi^{\text{out}}_{\mu\to i} \label{eq:rSP-g}\\
\Gamma_{0,\mu\to i} & =\frac{1}{m-1}\left(\partial_{\omega}^{2}\phi^{\text{out}}_{\mu\to i}-(2\partial_{V_1}\phi^{\text{out}}_{\mu\to i}-g_{\mu\to i}^{2})\right)\\
\Gamma_{1,\mu\to i} & =\frac{1}{m-1}(\partial_{\omega}^{2}\phi^{\text{out}}_{\mu\to i}-m(2\partial_{V_1}\phi^{\text{out}}_{\mu\to i}-g_{\mu\to i}^{2}))\\
\end{align}
Incoming messages on the input nodes are then given by
\begin{align}
B_{i\to\mu} & =\sum_{\nu\neq \mu}F_{i}^{\nu} g_{\nu \to i} \\
A_{0,i\to\mu} & =\sum_{\nu\neq \mu} (F_{i}^{\nu})^2\Gamma_{0,\nu\to i}\\
A_{1,i\to\mu} & =\sum_{\nu\neq \mu}(F_{i}^{\nu})^2 \Gamma_{1,\nu\to i} \label{eq:rSP-A1}\\
\end{align}

The closed set of Equations (\ref{eq:rSP-x}-\ref{eq:rSP-V1}) and (\ref{eq:rSP-g}-\ref{eq:rSP-A1}), along with the free entropy definitions in Eqs. \eqref{eq:phi_in_1RSB} and \eqref{eq:phi_out_1RSB}, define the rSP iterative message passing.

\subsection{The GASP Equations}
Under our statistical assumptions on the sensing matrix $\bF$, in order to reduce the computational complexity of rSP, it is possible to close the equations the rSP message passing in terms of single site or scalar quantities
$\omega_{\mu},g_{\mu}, \Gamma_{0,\mu}, \Gamma_{1,\mu}, A_{0}, A_{1}, B_{i}, \Delta_{0,i}, \Delta_{1,i}, V_{0}$ and $V_{1}$, therefore obtaining the GASP equation. In fact, the values $ A_{0,i\to \mu}, A_{1,i\to \mu}$ and $V_{0,\mu\to i}, V_{1,\mu\to i}$ concentrate and can be straightforwardly replaced by their scalar counterparts.
In order to present in this section all of the necessary ingredients of the GASP algorithm, we rewrite here the two scalar channel free entropies from previous section. 
Adopting a form that makes clear the nested structure of the 1RSB free-entropy and it's relation to the corresponding RS free  entropy used in GAMP, we write fro the input channel
\begin{align}
\phi^{\text{in}}(B,A_{0},A_{1},m)= & \frac{1}{m}\log\int Dz\ e^{m\varphi^{\text{in}}(B+\sqrt{A_{0}}z,\,A_{1})}\label{eq:phi-in-T}\\
\varphi^{\text{in}}(h , A_{1})= & \log\int_\chi \dd x\ e^{-\beta r(x)-\frac{1}{2}A_{1}x^{2}+h x}\label{eq:varphi-in-T}
\end{align}
and for the output channel
\begin{align}
\phi^{\text{out}}(\omega,V_{0},V_1,y,m)= & \frac{1}{m}\log\int Dz\ e^{m\varphi^{\text{out}}(\omega+\sqrt{V_{0}}\,z,\,V_1,y)}\label{eq:phi-out-T}, \\
\varphi^{\text{out}}(u, V_1, y)= & \log\int Dz\ e^{-\beta\ \ell(y,u+\sqrt{V_1}z )}\label{eq:varphi-out-T}.
\end{align}
As usual, $\int D z$ denotes standard Gaussian integration $\int dz\, \exp(-z^2/2)/\sqrt{2\pi}$.
We will use the notation $\phi^{\text{in}}_i=\phi^{\text{in}}(B_i,A_{0},A_{1},m)$ and
$\phi^{\text{out}}_\mu=\phi^{\text{out}}(\omega_\mu,V_{0},V_1,y_\mu,m)$ and drop time indexes for the time being.
Given the definition $B_i=\sum_\mu F^\mu_i g_{\mu\to i}$, we can write
\begin{align}
\hat{x}_{i\to\mu} & =\partial_{B}\phi^{\text{in}}(A_{0},A_{1},B_{i}-F_{i}^{\mu}g_{\mu\to i})\\
 & \approx\hat{x}_{i}-F_{i}^{\mu}g_{\mu}\partial_{B}^{2}\phi_{i}^{\text{in}},
\end{align}
which can be then inserted in the definition $\omega_\mu = \sum_i F^\mu_i \hat{x}_{i\to\mu}$ resulting in
\begin{equation}
\omega_{\mu}=\sum_{i}F_{i}^{\mu}\hat{x}_{i}-g_{\mu}\sum_{i}\left(F_{i}^{\mu}\right)^{2}\partial_{B}^{2}\phi_{i}^{\text{in}}.
\end{equation}
The other relevant equation is
\begin{align}
g_{\mu\to i} & =\partial_{\omega}\phi^{\text{out}}(V_{0},V_{1},\omega^{\mu}-F_{i}^{\mu}\hat{x}^{i\to\mu})\\
 & \approx g_{\mu}-F_{i}^{\mu}\hat{x}^{\mu}\partial_{\omega}^{2}\phi_{\mu}^{\text{out}},
\end{align}
which analogously leads to
\begin{equation}
B_{i}=\sum_{\mu}F_{i}^{\mu}g_{\mu}-\hat{x}_{i}\sum_{\mu}\left(F_{i}^{\mu}\right)^{2}\partial_{\omega}^{2}\phi_{\mu}^{\text{out}}.
\end{equation}
We now introduce back the time indexes, and use the shorthand notations $\phi^{\text{in},t}_{i}  =\phi^{\text{in}}(B^t_{i},A^t_{0},A^t_{1},m)$ and $\phi^{\text{out},t}_{\mu}= \phi^{\text{out}}(\omega_{\mu}^t,V^{t-1}_{0},V^{t-1}_1,y_{\mu},m)$. Using again the definition $c_{\bF}=\frac{1}{M N}\sum_{\mu,i} (F^\mu_i)^2$ (hence $\mathbb{E}\, c_{\bF} = 1/N$ in our setting), with some initialization for $\hat{x}^{t=0}_{i}$,$V^{t=0}_0,V^{t=0}_1$ and setting $g_{\mu}^{t=0}=0$, we finally obtain
\begin{align}
\omega^t_{\mu} & =\sum_{i}F_{i}^{\mu}\hat{x}^{t-1}_{i}-g^{t-1}_{\mu}(m V^{t-1}_{0}+V^{t-1}_{1}) \label{eq:gaspT-w}\\
g^t_{\mu}&=\partial_{\omega}\phi^{\text{out},t}_{\mu}\\
\Gamma^t_0 &= \frac{1}{m-1}\left(\partial_{\omega}^{2}\phi^{\text{out},t}_{\mu}-(2\partial_{V_1}\phi^{\text{out},t}_{\mu}-(g^t_{\mu})^{2})\right) \\
\Gamma^t_1 &= \frac{1}{m-1}(\partial_{\omega}^{2}\phi^{\text{out},t}_{\mu}-m(2\partial_{V_1}\phi^{\text{out},t}_{\mu}-(g^t_{\mu})^{2})) \\
A^t_0 & =c_{\bF}\sum_{\mu} \Gamma^t_0\\
A^t_1 & =c_{\bF}\sum_{\mu} \Gamma^t_1\\
B^t_{i} & =\sum_{\mu}F_{i}^{\mu}g^t_{\mu}-\hat{x}^{t-1}_{i}(m A^t_0-A^t_1)\\
\hat{x}^t_{i} & =\partial_{B}\phi_{i}^{\text{in},t} \label{eq:gaspT-x}\\
\Delta^t_{0,i} & =\frac{1}{m-1}\left(\partial_{B}^{2}\phi_{i}^{\text{in},t}+ 2\partial_{A_{1}}\phi_{i}^{\text{in},t}+(\hat{x}^t_{i})^{2} \right) \label{eq:gaspT-delta0}\\
\Delta^t_{1,i} & =\partial_{B}^{2}\phi_{i}^{\text{in},t}-m \Delta^t_{0,i}  \label{eq:gaspT-delta1}\\
V^t_{0} & =c_{\bF}\sum_{i}\Delta^t_{0,i}\label{eq:gaspT-V0}\\
V^t_{1} & =c_{\bF}\sum_{i}\Delta^t_{1,i}. \label{eq:gaspT-V1}
\end{align}
Equations (\ref{eq:gaspT-w}-\ref{eq:gaspT-V1}), along with the free entropy definitions in Eqs. (\ref{eq:phi-in-T}, \ref{eq:phi-out-T})  are the GASP iterative equations.

\subsection{Zero Temperature Limit}
In order to apply the GASP algorithm to MAP estimation, we have to consider the zero-temperature limit $\beta\uparrow\infty$ of the message passing. The limiting form of the equations depends on the model and on the regime (e.g. low or high $\alpha$). Here we consider models defined on continuous spaces $\chi^N$ and  in the high $\alpha$ regime (e.g. $\alpha >1$ for phase retrieval). In this case, while taking the limit, the messages have to be rescaled appropriately in order to keep them finite. Therefore, we rescale the messages through the substitutions
\begin{align}
A_{0} & \to\beta^{2}A_{0} \label{eq:scaling_in}\\
A_{1} & \to\beta A_{1}\\
B & \to\beta B\\
\omega & \to\omega\\
V_{0} & \to V_{0}\\
V_{1} & \to V_{1}/\beta\\
g & \to\beta g\\
m & \to m/\beta\\
\Delta_{0} & \to\Delta_{0}\\
\Delta_{1} & \to\Delta_{1}/\beta, \label{eq:scaling_fin}
\end{align}
in Equations (\ref{eq:gaspT-w}-\ref{eq:gaspT-V1}) and Eqs. (\ref{eq:phi-in-T}, \ref{eq:varphi-out-T}).
Taking the $\beta\to\infty$ limit we recover the GASP equations for MAP estimation presented in the Main Text.

\subsection{GASP equations for real-valued phase retrieval problem}

Putting together Eqs.\eqref{eq:varphi_in_PR} and \eqref{eq:varphi_out_PR}, and the definitions in Eqs.\eqref{eq:phi-in-T} and \eqref{eq:phi-out-T}, we
can obtain the zero temperature limit of the two GASP scalar estimation channels, 
in the special case of the phase retrieval loss $\ell(y,u)=(y-|u|)^2$ and an $L_2$-norm $r(x)=\lambda x^2/2$. The expressions simply become: 
\begin{align}
\phi^{\text{in}}(B,A_{0},A_1,y, m) = & -\frac{B^2}{2(A_1+\lambda-mA_0)} -\frac{1}{2m} \log\left(1-\frac{m A_0}{A_1+\lambda} \right)  \\
\phi^{\text{out}}(\omega,V_{0},V_1,m)= & \frac{1}{m}\log\left(Z_+ + Z_-\right) - \frac{1}{2m}\log\left(1+\frac{2 m V_0}{1+2 V_1}\right),
\end{align}
where we defined for compactness:
\begin{align}
Z_{\pm} = & H\left(-\frac{2 m V_0 y\mp \omega(1-2 V_1)}{\sqrt{V_0 (1+2 V_1)(1+2 V_1 +2 m V_0)}}\right) \exp\left(-\frac{m (\omega \pm y)^2}{1+ 2 V_1 +2 m V_0 } \right). 
\end{align}

Moreover, the zero temperature limit of GASP Eqs. (\ref{eq:gaspT-x}, \ref{eq:gaspT-delta0}, \ref{eq:gaspT-delta1}) after the rescaling discussed in previous paragraph, gives:
\begin{align}
\hat{x}_{i} & =\frac{B_{i}}{A_1+\lambda-mA_0}\\
\Delta_{i}^{0} & =\frac{A_0}{(A_1+\lambda)(A_1+\lambda-mA_0)}\\
\Delta_{i}^{1} & =\frac{1}{A_1+\lambda-mA_0}-m\Delta_{i}^{0}.
\end{align}

\section{Setting the symmetry-breaking parameter}

The 1RSB formalism, from which the (G)ASP equations are derived, is based on the introduction of a symmetry-breaking parameter, the so-called Parisi parameter $m$ \cite{mezard1987spin}, that allows the description of the fine structure of highly non-convex (``glassy'') landscapes.

In replica analyses, the physical meaning of $m$ is the following: when the studied model develops a 1RSB structure, by tuning $m$ in its natural range of variability $(0,1]$, it is possible to focus the Gibbs measure on the different families of exponentially numerous ``states'' (i.e., basins of solutions of the inference problem) that populate the loss landscape \cite{mezard1987spin}. 
The dominant states, i.e. those where a perfect sampling algorithm would land with high probability, are described at the thermodinamically optimal value $m^{\star}$, that extremizes the free-energy of the model.

In the real-replica formalism employed to derive the ASP equations \cite{monasson1995structural,antenucci2019approximate}, however, $m$ is an external parameter that can be analytically continued to take any real value, and is no-longer strictly bound to the interval $(0,1]$. In fact, both the algorithm and its SE characterization are valid even if the model has not developed a proper 1RSB structure, and $m$ can be simply thought as a parametrization the family of algorithms ASP$(m)$ \cite{antenucci2019approximate}. We note that, in the zero-temperature limit, when the proper scaling of $m$ with $\beta\to\infty$ is chosen (Eqs.~\ref{eq:scaling_in} to \ref{eq:scaling_fin}), even the physically meaningful interval of variability of $m$ is of course extended to $(0,\infty)$.

\begin{figure}[ht]
\begin{center}
\includegraphics[width=0.75\textwidth]{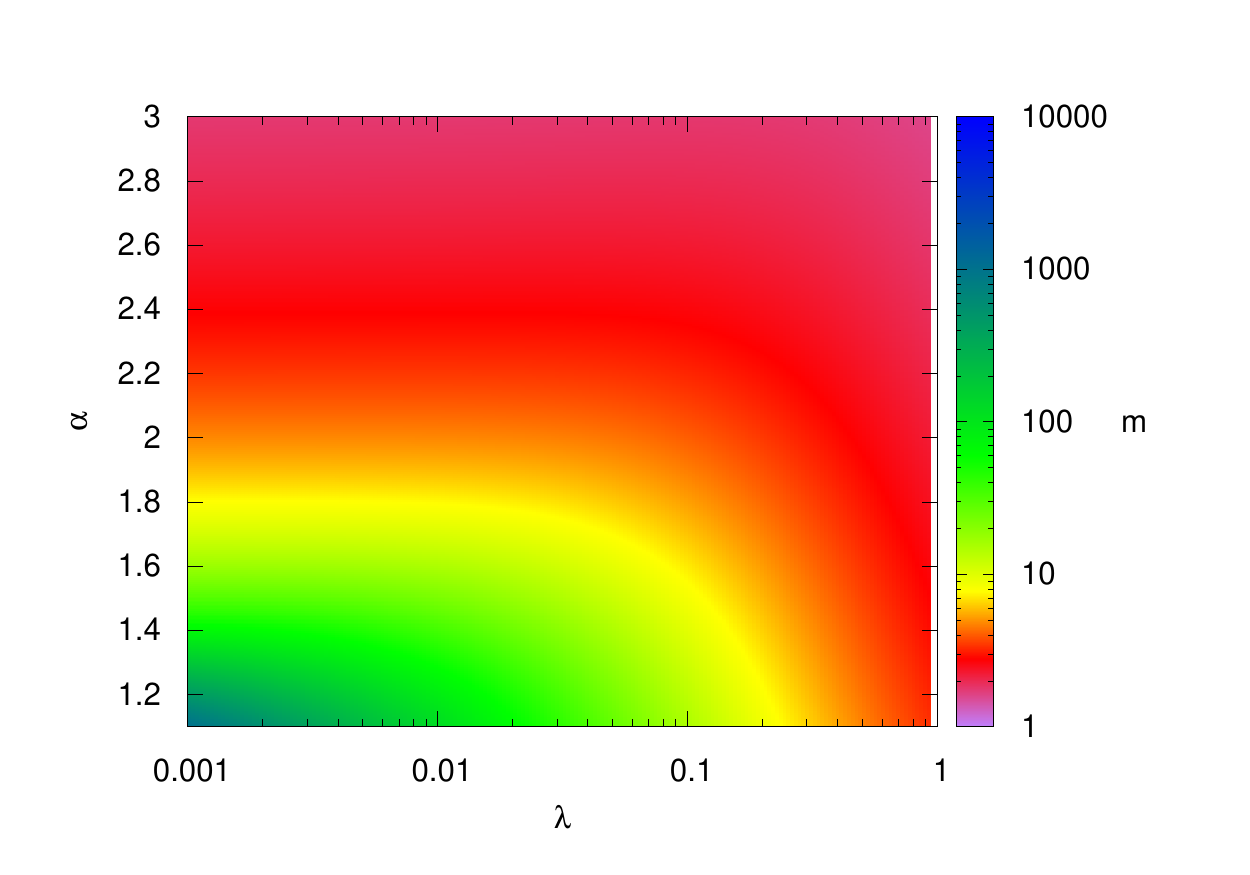}
\end{center}
\caption{\label{fig:m_values} Optimal value of the symmetry-breaking parameter $m=m^{\star}$ (as employed in the GASP phase diagram in Fig. 2 in the Main Text, bottom plot), for different values of the regularizer $\lambda$.}
\end{figure}

In Fig.~\ref{fig:m_values}, we show the numerical values of the thermodynamic optima $m=m^{\star}$ in the zero-temperature phase retrieval problem (obtained analytically in correspondence of $\rho=0$, at varying values of $\alpha$ and $\lambda$, from a replica computation that will be presented in a more technical future work). These are the values that were employed in the corresponding GASP phase diagram, presented in the Main Text in Fig. 2. 

We remark, however, that this particular choice was mostly due to the need of consistency in the criterion for fixing $m$ throughout the various regions of the phase diagram. In fact, as it was already noted in the Bayesian case \cite{antenucci2019approximate}, the thermodynamical optimum might not be the best choice for $m$, since other values seem to allow better inference (e.g., a decreased final MSE). Since we are here interested in the MAP estimation task, our performance evaluation is based solely on the possibility of achieving retrieval of the signal. This condition is definitely less demanding than that of obtaining the best MSE, and in fact we find that wide ranges of values for $m$ are effective in correspondence of each $\alpha$ and $\lambda$. Fig.~\ref{fig:m_values} is nevertheless indicative of how $m$ should be incremented when the observation matrix gets smaller or when weaker regularizers are employed.    

\begin{figure}[ht]
\begin{center}
\includegraphics[width=0.75\textwidth]{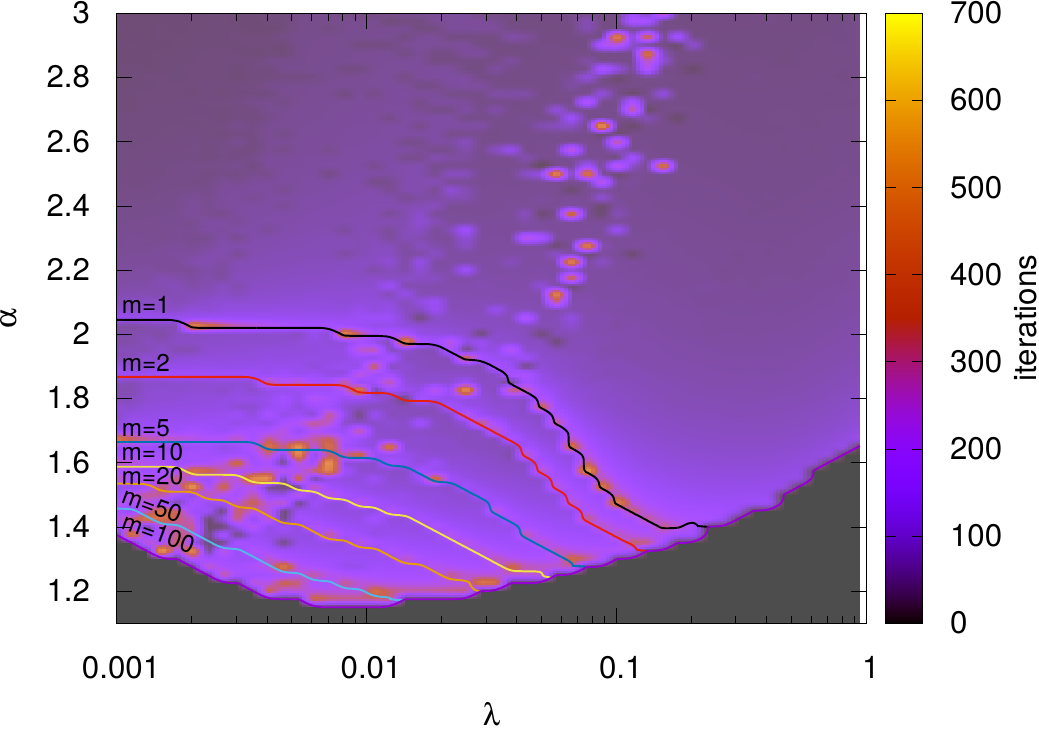}
\end{center}
\caption{\label{fig:fixed_m} Total number of iterations to convergence for  GASP$(m)$. The colored curves delimit (from below) the perfect recovery regions of GASP with the indicated value for $m$. }
\end{figure}
In order to show the robustness of GASP$(m)$ with respect to the choice of different values for the symmetry-breaking parameter, in Fig.~\ref{fig:fixed_m} we plot the total number of iterations required to converge to the signal (indicated by the color map), for fixed values of $m$. The plotted number of iterations include both stages in our simple \emph{continuation} strategy. As it can be seen in the plot, this total number tends to increase as $\alpha$ is lowered, since the inference problem becomes harder.

The colored curves mark the lower border of the regions of effectiveness of GASP$(m)$, with $m$ fixed in each region, at which the number of iterations required by the algorithm diverge. It is clear, indeed, that a careful fine-tuning of $m$ is unnecessary, and that it is quite intuitive how to adapt it when a different instance of the problem is given. For example, in the noiseless case, a basic strategy is to fix $\lambda$ in the range $[0.001:0.01]$ and then test $\mathcal{O}(1)$ different values for $m$, until $MSE=0$ is obtained at convergence of the message-passing.

\begin{figure}
\includegraphics[width=\columnwidth]{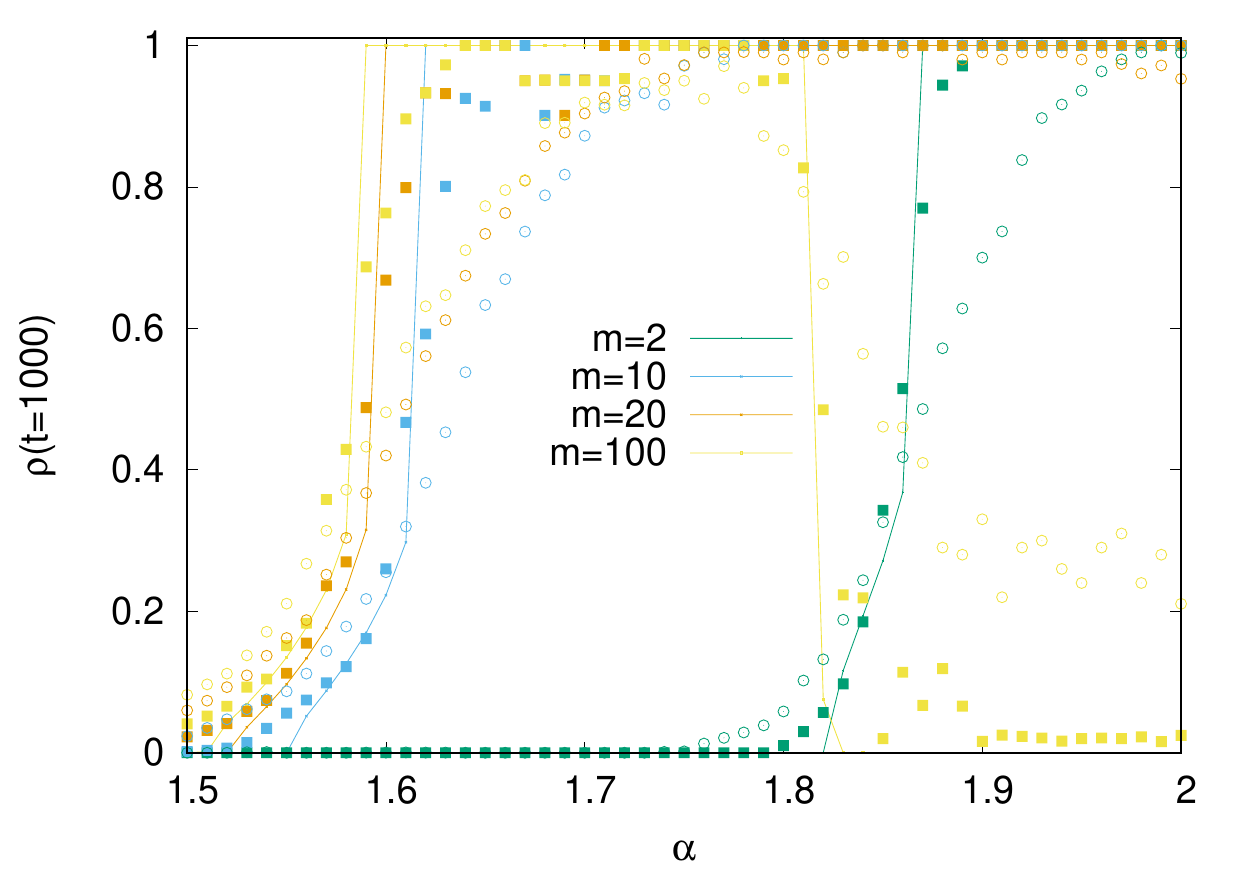}
\caption{GASP and SE result after $t=10^3$ iterations. Start at $t=0$ with $\rho=10^{-3}$ for SE and $\hat{\bx}\sim\mathcal{N}(0,I_N)$ for GASP. Circles are for $N=10^3$, squares for $N=10^4$, . results are averaged over $100$ samples. Lines are predictions from SE.}
\label{fig:scaling}
\end{figure}

As a last data point, we report in Fig. \ref{fig:scaling} the behaviour of the overlap with the true signal of the estimator given by GASP, for two different system sizes, large times and as a function of $\alpha$. We observe that for large $N$ transitions become sharper and experimental points approach the asymptotic prediction from SE.

\end{document}